\documentclass[a4paper,fleqn,usenatbib,useAMS,usedcolumn]{mnras}

\usepackage[T1]{fontenc}
\usepackage{ae,aecompl}
\usepackage[utf8]{inputenc}

\usepackage{graphicx} 
\DeclareGraphicsExtensions{.pdf,.png,.jpg,.ps}
\graphicspath{{./plots/}}

\usepackage{amsmath} 
\usepackage{amssymb}  
\usepackage{gensymb}
\usepackage{hyperref}
\hypersetup{
    colorlinks = true,
    citecolor ={MidnightBlue}
}
\usepackage[nolist]{acronym}
\usepackage{ulem}
\usepackage{bm}
\usepackage{etoolbox}
\usepackage[inline]{enumitem}	
\usepackage[dvipsnames]{xcolor}

\definecolor{mustard}{rgb}{1.0, 0.86, 0.35}
\definecolor{cyan(process)}{rgb}{0.0, 0.72, 0.92}
\definecolor{ochre}{rgb}{0.8, 0.47, 0.13}
\definecolor{plum}{HTML}{5A1E5C}

\acrodef{BNS}{binary neutron star}
\acrodefplural{BNS}[BNSs]{binary neutron stars}
\acrodef{DCO}{double compact object}
\acrodefplural{DCO}[DCOs]{double compact objects}
\acrodef{NS}{neutron star}
\acrodefplural{NS}[NSs]{neutron stars}
\acrodef{GRB}{gamma--ray burst}
\acrodef{RLOF}{Roche lobe overflow}
\acrodef{CE}{common envelope}
\acrodef{RL}{Roche lobe}
\acrodefplural{RL}[RLs]{Roche lobes}
\acrodef{HeS}{Helium star}
\acrodef{SN}{supernova}
\acrodefplural{SN}[SNe]{supernovae}

\title[Stellar response after stripping]
{Stellar response after stripping as a model for common-envelope outcomes}

\author[]{\parbox{\textwidth}{Alejandro Vigna-G\'{o}mez$^{1,2,3,4}$\thanks{E-mail: avignagomez@nbi.ku.dk}, Michelle Wassink$^{5,3}$, Jakub Klencki$^{5}$,  Alina Istrate$^{5}$, Gijs Nelemans$^{5,6,7}$ and
Ilya Mandel$^{3,4,8}$}
\vspace{0.5cm}\\
\parbox{\textwidth}{
$^{1}$DARK, Niels Bohr Institute, University of Copenhagen, Jagtvej 128, 2200, Copenhagen, Denmark\\
$^{2}$Niels Bohr International Academy, The Niels Bohr Institute, Blegdamsvej 17, 2100 Copenhagen, Denmark\\
$^{3}$Monash Centre for Astrophysics, School of Physics and Astronomy, Monash University, Clayton, Victoria 3800, Australia\\
$^{4}$The ARC Center of Excellence for Gravitational Wave Discovery -- OzGrav, Australia\\
$^{5}$Department of Astrophysics/IMAPP, Radboud University, P.O. Box 9010, 6500 GL Nijmegen, The Netherlands\\
$^{6}$Institute of Astronomy, KU Leuven, Celestijnenlaan 200D, B-3001 Leuven, Belgium\\
$^{7}$SRON, Netherlands Institute for Space Research, Sorbonnelaan 2, NL-3584 CA Utrecht, The Netherlands\\
$^{8}$Birmingham Institute for Gravitational Wave Astronomy and School of Physics and Astronomy, University of Birmingham, Birmingham, B15 2TT, United Kingdom
}}

\date{\today}

\pubyear{2021}

\begin{document}

\label{firstpage}
\pagerange{\pageref{firstpage}--\pageref{lastpage}}
\maketitle

\begin{abstract}
Binary neutron stars have been observed as millisecond pulsars, gravitational-wave sources, and as the progenitors of short gamma-ray bursts and kilonovae. 
Massive stellar binaries that evolve into merging double neutron stars are believed to experience a common-envelope episode. 
During this episode, the envelope of a giant star engulfs the whole binary.
The energy transferred from the orbit to the envelope by drag forces or from other energy sources can eject the envelope from the binary system, leading to a stripped short-period binary. 
In this paper, we use one-dimensional single stellar evolution to explore the final stages of the common-envelope phase in progenitors of neutron star binaries.
We consider an instantaneously stripped donor star as a proxy for the common-envelope phase and study the star's subsequent radial evolution. 
We determine a range of stripping boundaries which allow the star to avoid significant rapid re-expansion and which thus represent plausible boundaries for the termination of the common-envelope episode.  
We find that these boundaries lie above the maximum compression point, a commonly used location of the core/envelope boundary. 
We conclude that stars may retain fractions of a solar mass of hydrogen-rich material even after the common-envelope episode.
If we consider orbital energy as the only energy source available, all of our models would overfill their Roche lobe after ejecting the envelope, whose binding energy includes gravitational, thermal, radiation, and recombination energy terms.

\end{abstract}


\section{Introduction}
\label{sec:Introduction}
In the last decades, close binaries have been detected in the Milky Way as  cataclysmic variables \citep[e.g.,][]{1987A&AS...70..335R,1988QJRAS..29....1K,2011ApJS..194...28K}, X-ray systems \citep[e.g.,][]{THstandaardscenario,2011Ap&SS.332....1R,2011Natur.479..372K}, spectroscopic binaries \citep[e.g.,][]{1988ApJ...334..947S} and pulsar binaries \citep[e.g.,][]{HulseTaylor,2013Sci...340..448A}.
Late stages of massive binary evolution, such as the merger of \ac{NS} binaries, have been long speculated as the progenitors of short gamma-ray bursts \citep{1986ApJ...308L..43P}.
Galactic \acp{BNS}, which are typically observed as recycled radio pulsars \citep[][and references therein]{TaurisFormation}, have shown that very close binaries are driven by gravitational-wave emission toward coalescence \citep{2005ASPC..328...25W}.
On August 17, 2017, the coincident detection of a gravitational-wave signal, a short gamma-ray burst, and a kilonova \citep{FirstDNSLigo,Multimessenger}, successfully confirmed the theory of \ac{BNS} mergers as gravitational-wave sources and their electromagnetic counterparts. 

Apart from GW170817 \citep{FirstDNSLigo} and GW190425 \citep{2020ApJ...892L...3A}, the first and second gravitational-wave signals from \ac{BNS} mergers, respectively, the LIGO and Virgo scientific collaborations have detected the coalescence of tens of binary black holes and two \ac{NS} -- black hole binaries throughout their three observing runs \citep{LigoO1+2,2021PhRvX..11b1053A,GW200105,2021arXiv211103606T}.
While merging binary black holes can be assembled through dynamical processes in dense stellar environments \citep[e.g.,][]{Dynamicalfor3, Dynamicalfor, Dynamicalfor2}, these are not expected to significantly contribute to the population of merging \ac{BNS} \citep[e.g.,][]{2020ApJ...888L..10Y}, which is expected to be dominated by isolated binaries in the field \citep[e.g.,][]{Dynamicalun, FirstNSNS}.  
In order to form merging double compact objects through isolated binary evolution, binaries typically need to start at sufficiently wide separation ($\sim 10^2-10^4\ \rm{R_{\odot}}$) to avoid early stellar mergers, but must yield close ($\sim 0.1-10\ \rm{R_{\odot}}$) double compact objects in order to be able to merge within the age of the Universe (see, e.g., \citealt{MandelFarmer:2018,2021hgwa.bookE...4M} for recent reviews).    
Therefore, the transition from a wide stellar binary to a close double compact object is arguably the main problem in the theory of \ac{BNS} formation.

One of the most studied mechanisms for the formation of close binaries is via the \ac{CE} phase \citep[e.g.,][]{PacCE,vdHeuvel1976,IvanovaReview}. 
The \ac{CE} is the outcome of a dynamically unstable mass transfer episode. 
At the onset of the dynamical instability, the gaseous envelope of the donor star engulfs the companion and loses co-rotation with the binary composed of the donor core and the companion.  The drag forces between this binary and the \ac{CE} dissipate orbital energy and cause the binary to spiral in. 
The possible outcomes of a \ac{CE} phase are either a successful ejection of the envelope, which results in a close binary, or a merger. 
The \ac{CE} phase is a complex process.
It is particularly challenging to model numerically because it involves physics from different spatial and temporal scales \citep[][and references therein]{IvanovaReview}.
To avoid these difficulties, the \ac{CE} phase has been frequently parameterised by the so-called energy formalism in binary population synthesis studies \citep{Webbinkalpha}. 

In the energy formalism, the orbital outcome of the \ac{CE} phase is predicted by comparing the change in the orbital energy ($\Delta E_{\textrm{orb}}$) with the gravitational binding energy of the envelope  \begin{equation}
\label{eqn:bindingEnergy}
E_{\textrm{bind}} = \int_{\textrm{core}}^{\textrm{surface}} (\Psi(m)+\epsilon(m))dm =  -\frac{GMM_{\textrm{env}}}{\lambda R},
\end{equation}
where $\Psi$ is the gravitational potential, $\epsilon$ is the specific internal energy, $G$ is the gravitational constant, $M$ is the total mass of the star, $M_{\textrm{env}}$ is the mass of the stellar envelope, $R$ is the stellar radius, and $\lambda$ is a dimensionless parameter that depends on the structure of the star \citep{Koollambda}.
This comparison is usually done by introducing a dimensionless efficiency parameter $\alpha$ such that 
$\alpha \times \Delta E_{\textrm{orb}} = E_{\textrm{bind}}$, which can be rewritten as
\begin{equation}
\label{eqn:orbitaComplete}
   \alpha \Big( \frac{GM_{\textrm{don,rem}} M_{\textrm{acc}}}{2a_{\textrm{f}}} - \frac{GM_{\textrm{don}}M_{\textrm{acc}}}{2a_{\textrm{i}}} \Big) = \frac{GM_{\textrm{don}}M_{\textrm{don,env}}}{\lambda R_{\textrm{don}}},
\end{equation}
where the \textit{don} and \textit{acc} subscripts refer to donor and accretor, respectively, $M_{\textrm{rem}}$ is the mass of the core remnant after the CE, $a$ is the semi-major axis, and the $i$ and $f$ subscripts refer to the initial and final states of the \ac{CE} event, respectively.

Equation~\ref{eqn:orbitaComplete}, sometimes referred to as the $\alpha \lambda$ formalism, is arguably the most used formula in the \ac{CE} literature.
However, the parameterisation in the energy formalism is a major source of uncertainty.
The efficiency of transferring gravitational energy from the inspiral to the common envelope has been thoroughly discussed in the literature \citep[e.g.,][]{IvanovaReview}, and the interpretation usually varies.
In the most simple case, when the \ac{CE} episode is considered as a purely gravitational problem, the efficiency parameter takes values between $0 < \alpha \leq 1$.
The $\alpha \ll 1$ regime corresponds to the case when most of the energy is lost from the system, which results in extreme inspirals, while $\alpha=1$ results in a perfectly efficient energy transference and an apparent upper limit on binary hardening.
However, additional physical processes, such as accretion feedback or nuclear burning, can lead to $\alpha>1$ \citep{Podsiadlowski2010,IvanovaCEPitfalls,Ivanova2015}.
Despite significant recent efforts to understand and quantify the $\alpha$ efficiency parameter for massive donors \citep[e.g.,][]{FragosMesa,2020arXiv201106630L}, it remains very uncertain.

Another source of uncertainty lies in the parameterization of the binding energy $E_{\textrm{bind}}$ \citep{IvanovaCEPitfalls}.
The calculation of the binding energy of the envelope is frequently simplified by using the $\lambda$ parameter as defined in Eq.~\ref{eqn:bindingEnergy}.
The $\lambda$ parameter can be determined by using detailed (but one-dimentional) stellar models, and its value, which can span orders of magnitude, will depend strongly on the envelope type (radiative or convective) and the evolutionary stage of the donor star \citep{1994MNRAS.270..121H,DewiTauris2000, PodsiadBHB, 2010ApJ...716..114X,2010ApJ...722.1985X, KkruckowCEejection, Wangformationlambda, Klencki2021A&A}.
This parameter becomes the sole container of all the details about the structure of the star.

Importantly, these calculations rely on the assumed boundary between the ejected envelope and the remaining core, i.e. the lower limit of the integral in Eq.~\ref{eqn:bindingEnergy}, usually referred to as the \textit{bifurcation} or the \textit{divergence} point \citep{DewiTauris2000,Tauris2001,IvanovaCEFate,deMarco2011,KkruckowCEejection}. It is thought to be located somewhere between the hydrogen-depleted core and the bottom of the convective envelope. \citet{KkruckowCEejection} showed that the exact location of the bifurcation point can influence the binding energy of convective-envelope stars by more than two orders of magnitude, making this is a crucial issue.

There is no unequivocal definition of the bifurcation point location
\citep[e.g.,][]{deMarco2011,2021A&A...650A.107M}.
\citet{IvanovaCEFate} argued for the point of maximum compression in the H-burning shell, i.e., $M_{\textrm{cp}} \equiv \max_{\textrm{local}}(P/\rho)$, to avoid any immediate expansion of the donor remnant after the CE ejection.
Others have placed the bifurcation point at the location in the envelope where the hydrogen abundance is $X = 0.1$ \citep[e.g.,][]{KkruckowCEejection,Klencki2021A&A}. 
It has also been defined at the coordinate of hydrogen exhaustion, maximum energy generation, $X=0.15$, and the lower boundary of the convective envelope, to name just a few choices \citep[see][and references therein]{IvanovaCEPitfalls}.
These choices can lead to markedly different outcomes when computing the binding energy and applying Eq.~\ref{eqn:orbitaComplete}.

The bifurcation point also determines the amount of mass removed during the \ac{CE} phase and this in turn affects the stellar composition of the stripped star and its subsequent evolution, including the amount of mass loss.
A bifurcation point in hydrogen depleted layers leads to a fully stripped star that becomes a compact helium star with strong stellar winds, the likely progenitors of type Ib/c supernovae \citep{TaurisUSS}.
Partially stripped stars retain some hydrogen in the envelope \citep{2017A&A...608A..11G}, which can lead to significant subsequent re-expansion \citep{Laplace2020L} and result in (hydrogen-rich) type II supernovae.
Envelope expansion is particularly relevant for systems with a pulsar companion, as mass transfer can spin up and recycle the pulsar \citep[e.g,][]{TaurisFormation}, as seen in the observed millisecond pulsars in \acp{BNS} in the Galaxy. However, this can also be due to a later phase of expansion of fully stripped stars \citep[e.g.,][]{TaurisFormation,VignaGomez2018}.

In this paper, we investigate the bifurcation point by considering the response of a single star to rapid stripping as a proxy for the common envelope evolution (see also \citealt{IvanovaCEFate,HallTout:2014,2018MNRAS.480.5176H}).  We  consider a range of possible stripping boundaries.  If the star rapidly expands beyond the stripping boundary, it is unlikely to emerge from the common-envelope phase in that configuration, and would likely be stripped further.  On the other hand, a star which rapidly contracts after stripping indicates excessive mass loss during the common-envelope phase.  This suggests that the bifurcation point, and a stable post-common-envelope configuration, lies between these two regimes.

In Sec.~\ref{sec:methods}, we present the method to perform this study.
In Sec.~\ref{sec:results}, we present the results of how varying the amount of stripped mass affects the prompt radial re-expansion of a star and thus determine the range of possible bifurcation points and explore the final post-\ac{CE} separation for these cases; we do this for different donor masses and metallicities. 
In Sec.~\ref{sec:discussion}, we discuss the implication of our results in the context of binary evolution, particularly in the formation of \acp{BNS}, and present our concluding remarks.

\section{Numerical methods}
\label{sec:methods}
 We use the MESA stellar evolution code \citep[][version r11554]{Paxton2011, Paxton2013, Paxton2015, Paxton2018, Paxton2019} to simulate the evolution  of a massive star before and after the \ac{CE} phase.  
The numerical setup is divided in three different stages as following: (i) we evolve a massive star, of a given mass and metallicity, from the pre-main-sequence until it reaches a stellar radius of 500~R$_{\odot}$ (Sec.~ \ref{sub:progenitor}), then (ii) we mimic the \ac{CE} itself by artificially removing, partially or fully, the hydrogen-rich envelope, and then relaxing the remnant model (Sec.~\ref{sub:stripping}), and finally (iii) we further evolve the stripped star to probe the radial evolution until later stages of nuclear burning (Sec.~\ref{sub:poststripping}).
We use the post-stripping properties of the star to estimate whether this degree of stripping would be consistent with surviving the common envelope (Sec.~\ref{sub:ejection}).
Finally, we present how we calculate the binding energy parameter $\lambda$ and define a post-\ac{CE} Roche-filling factor $f_R$ (Sec.~\ref{sub:lambdaCalculation}).

\subsection{Progenitor models}
\label{sub:progenitor}
We compute our models starting from the pre-main-sequence stage.
We consider progenitors with metallicities $Z=Z_{\rm{LMC}}=0.0047$ \citep{BrottRotating}, $Z=Z_{\rm{MW}}=0.0088$ \citep{BrottRotating}, and $Z=Z_{\rm{\odot}}=0.017$ \citep{Grevesse1996}, representatives of the average composition in the Large Magellanic Cloud (LMC), Milky Way (MW), and the Sun, respectively. 
We consider stars with initial masses $M=\{10,12,15,17,20,25\}\ \rm{M_{\odot}}$.

We use the \texttt{approx21.net} nuclear network, which includes reactions relevant for hydrogen burning through the CNO cycle and non-explosive helium burning up until alpha chains \citep{Paxton2011}. 
We use mixing length theory \citep{MLTHenyey} to account for convective mixing, with a mixing length parameter $\alpha_{\textrm{MLT}} = 1.5$. We apply the Ledoux criteria with the semiconvection parameter set to $\alpha_{\textrm{sem}} = 1$  \citep{semiconvection}. 
During the main sequence, we allow step overshooting above the convective hydrogen-burning core and up to a pressure scale height of 0.335 \citep{BrottRotating}.
We do not allow overshooting above convective zones in the shell, as those are not well constrained. 
We also included thermohaline mixing \citep{thermohaline}.
We modeled stellar winds following \cite{BrottRotating}. 
Our stars are not rotating. 

\begin{figure}
\includegraphics[width=\columnwidth]{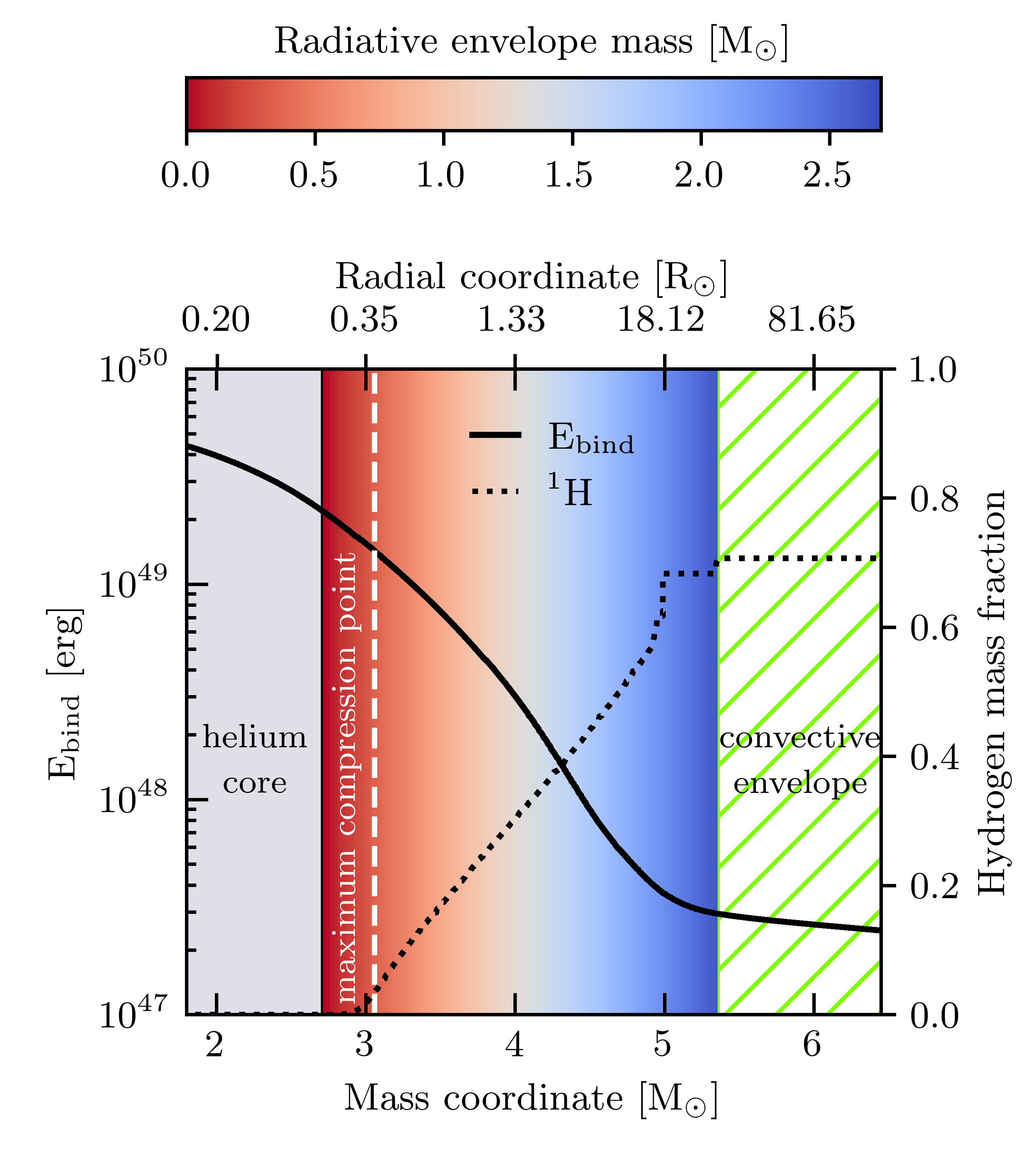}
\caption{
The internal profile of a star with initial mass of 12 $\rm M_{\odot}$ ($Z = Z_{\odot}$) at the onset of the \ac{CE} phase as a function of mass (bottom x-axis) and radial (top x-axis) coordinates.
At this point, the star has a total mass of 11.60\ $\rm M_{\odot}$, a helium core mass of 2.71\ $\rm M_{\odot}$, and a radius of $R \approx 500\  \rm R_{\odot}$, which extends beyond the convective region portrayed here.
The helium core is shown in grey,
the mass of the envelope in the hydrogen-shell region is shown as a red-to-blue coloured gradient (see colour bar), and the convective envelope is displayed as a green hatched region. We show the gravitational binding energy as a solid black curve (left y-axis) and the hydrogen abundance as a dotted black curve (right y-axis). The maximum compression point, which is sometimes chosen as the bifurcation point, is shown as a dashed white vertical line.  
}
\label{fig:StarProfile12Zsun}
\end{figure}

We stop the evolution of our stellar models when their radii reach $R = 500\ \rm{R_{\odot}}$, assuming that is the onset of the \ac{CE} phase.
At this radius, the lighter models ($M \leq 15\ \rm{M_{\odot}}$) are on the red giant branch and have an outer convective envelope, while the more massive models ($M \geq 17\ \rm{M_{\odot}}$) are still crossing the Hertzsprung gap and have a radiative outer envelope. 
For illustration purposes, in Fig.~\ref{fig:StarProfile12Zsun} we show the internal profile of a  12 M$_{\odot}$ progenitor with $Z=Z_{\odot}$ at  the onset of the CE phase, i.e. $R=500\ \rm{R_{\odot}}$.

\subsection{Stripping: emulation of the CE phase}
\label{sub:stripping}
We emulate the mass loss during \ac{CE} evolution by artificially stripping the progenitor star through  post-processing. This method allows for an easy and quick way to construct stripped models with different envelope masses.
An alternative to this stripping method can be, e.g., to induce a high, adiabatic, wind mass-loss rate of $1\ \rm M_{\odot}/yr$ \citep{IvanovaCEFate}; however, MESA struggles to numerically converge with such high mass-loss rates.   
The post-\ac{CE} star is constructed as follows.  From the MESA model, we extract the composition and the entropy profile of the donor star at the onset of the \ac{CE}, defined here as the moment when the star reaches a radius of 500 R$_{\odot}$. We define the core boundary as the central region where $X_H\approx10^{-6}$, and we integrate mass from the core boundary towards the surface until we reach the target envelope mass. All the remaining mass above this point is then discarded. We  use this composition  and entropy profile  to create the post-\ac{CE} star using the \texttt{relax\_initial\_composition} and \texttt{relax\_initial\_entropy}  methods available in MESA.  

The use of a pre-interaction stellar profile ignores both the redistribution of energy in the envelope prior to the dynamical inspiral phase and energy deposition beneath the stripping radius during the dynamical inspiral phase.  While both of these assumptions merit further investigation, they are commonly used in detailed hydrodynamical simulations of the common envelope \citep[e.g.,][]{2020arXiv201106630L}.
 
\subsection{Post-stripping evolution}
\label{sub:poststripping}
After the star has been stripped to emulate the \ac{CE} phase,  we proceed to evolve the post-\ac{CE} remnant  using MESA.
The numerical choices remain the same as during the pre-\ac{CE} evolution.
We evolve the stripped remnants for $2\times 10^5$ years, which is sufficient for all remnants to enter the core-helium burning stage and regain thermal equilibrium. 

\subsection{Emerging from a CE phase}
\label{sub:ejection}
We evaluate whether or not the \ac{CE} ejection could be successful by considering the radial re-expansion of the stripped remnant immediately after the emulated \ac{CE} phase. 

We assume that at the end of the \ac{CE} phase, the remnant approximately fills its Roche lobe. In response to the rapid envelope loss during the dynamical \ac{CE} phase, a stripped star will adjust its structure (and radius) in order to regain thermal equilibrium. For the stars considered in this study, this readjustment (referred to as the \textit{thermal pulse} by \citealt{IvanovaCEFate}) takes several hundred years. During this time, the stripped remnant can expand and immediately overflow its Roche lobe, thus effectively prolonging the \ac{CE} phase as well as the inspiral: we do not consider such models experienced a successful \ac{CE} ejection and consider them as potential mergers. 

The degree of expansion of the remnant will depend on the amount of remaining envelope left on top of the helium core (i.e. the guess at the bifurcation point location).  Larger remaining envelope mass leads to a more significant immediate expansion of the stripped remnant \citep{IvanovaCEFate}.  Here, we make the \textit{ad hoc} definitions that ``immediate'' is within 1000 yr and ``significant'' is by more than 5\%.  In other words, we assume that a successful \ac{CE} ejection is possible for a given remaining envelope mass if the remnant does not expand by more than 5\% within the 1000 yr following the end of the \ac{CE} inspiral.\footnote{As we will find out, some stripped remnants will re-expand on a longer timescale of $\sim10^4$ yr. We assume that this would lead to a separate Roche-lobe event rather than prolonging the \ac{CE} phase.}  The bifurcation point then refers to the maximum amount of mass that can remain on the donor while satisfying the successful \ac{CE} ejection condition.

The key moment in time in the above considerations is the end of the \ac{CE} phase. Our stripping method is essentially instantaneous. In reality, a \ac{CE} inspiral has a finite and largely uncertain duration \citep{IvanovaReview}. To take this into account, we assume that the initial stage of the post-stripping simulation takes place within the \ac{CE} phase.
We consider two possibilities for the \ac{CE} duration: a \textit{short} and a \textit{long} \ac{CE} phase. The short \ac{CE} phase, $t_{100}$, ends 100 yr after we restart the evolution of the stripped model.
The long \ac{CE} phase, $t_{1000}$, ends 1000 yr after we restart the evolution of the stripped model.
In this way, we define \textit{lower} and \textit{upper} bifurcation points, i.e. the remnant does no expand by more the 5 per cent within 1000 years from the end of the short ($t_{100}$) and long ($t_{1000}$) \ac{CE}, respectively. 
When computing quantities such as the post-\ac{CE} radius of the stripped donor and the associated amount of orbital inspiral, we use the stellar profiles at either $t_{100}$ or $t_{1000}$.

In addition to the criteria for the bifurcation point described above, we also consider \ac{CE} outcomes in which the donor is stripped down to the maximum compression point, sometimes assumed in the literature (see Sec.~\ref{sec:Introduction} and references therein). In that case, we assume 100 yr for the duration of the \ac{CE} phase 
and use the stellar profiles at $t_{100}$ to obtain post-\ac{CE} properties of the remnant. 

\subsection{Calculation of \texorpdfstring{$\lambda$}{1} and an estimate of the nominal post-CE Roche-filling factor \texorpdfstring{$f_R$}{1}}
\label{sub:lambdaCalculation}
In order to contextualize our results and compare them with others in the literature, we estimate the value of $\lambda$ for the critical points: the maximum compression, upper, and lower bifurcation points. 
We calculate the binding energy of the giant star models including gravitational and internal energy, and solve Eq.~\ref{eqn:bindingEnergy}, using the stripping point as the core/envelope boundary and the pre-\ac{CE} stellar profile to integrate through the envelope.
The internal energy includes the thermal energy of the gas, the energy of radiation, recombination energy, and dissociation energy.

Additionally, we define a parameter to estimate the nominal post-\ac{CE} Roche-filling factor
\begin{equation}
\label{eqn:fR}
    f_R:=\dfrac{\max|_{t<2000\ \rm{yr}}(R_\mathrm{stripped})}{R_{\rm{RL,f}}},
\end{equation}
where $R_{\rm{RL,f}}(a_{\rm{f}},q_{\rm{f}} = M_{\rm{don,rem}} / M_{\rm{acc}})$ is the Roche radius after stripping, which we approximate following \cite{EggletonRL} in the form
\begin{equation}
    R_{\rm{RL}}(a,q) = \frac{0.49}{0.6 + q^{-2/3} \ln(1 + q^{1/3})}a.
\end{equation} 
To calculate the post-\ac{CE} Roche radius, we consider that the orbital energy is 100\% efficient in unbinding the envelope ($\alpha=1$) and solve for the final separation using Eq. \ref{eqn:orbitaComplete}.
We consider the case when the companion is a \ac{NS} with mass $M_{\rm{acc}}=1.4\ \rm {M_{\odot}}$, and the donor is barely filling its Roche lobe at the onset of the \ac{CE} phase, i.e., $R_{\rm{RL,i}}(a_{\rm{i}},q_{\rm{i}} = M_{\rm{don}} / M_{\rm{acc}})=500\ \rm{R_{\odot}}$.
For values of $f_R>1$, we assume the stripped star will likely fill its Roche lobe at some point shortly ($t<2000$ yr) after the end of the \ac{CE} phase, resulting in an additional mass transfer episode and, potentially, a merger.
This approach does not track the orbital evolution during or after the common-envelope phase.

\section{Results}
\label{sec:results}
Here we present the main results of our stellar models.
First, we focus on the details of the radial post-stripping evolution and structure of two illustrative models (Sec.~\ref{sub:radius}).
Then, we summarise the results and show the main trends with varying progenitor masses and metallicities (Sec.~\ref{sub:comparison}).
Finally, we use our results to estimate and report the values of $\lambda$, $f_R$, and the bifurcation points (Sec.~\ref{sub:alphalambdaresult}).
Finally, we quantify the effect of the bifurcation point in the late stages of the \ac{CE} phase.

\begin{figure}
\includegraphics[width=1.0\columnwidth]{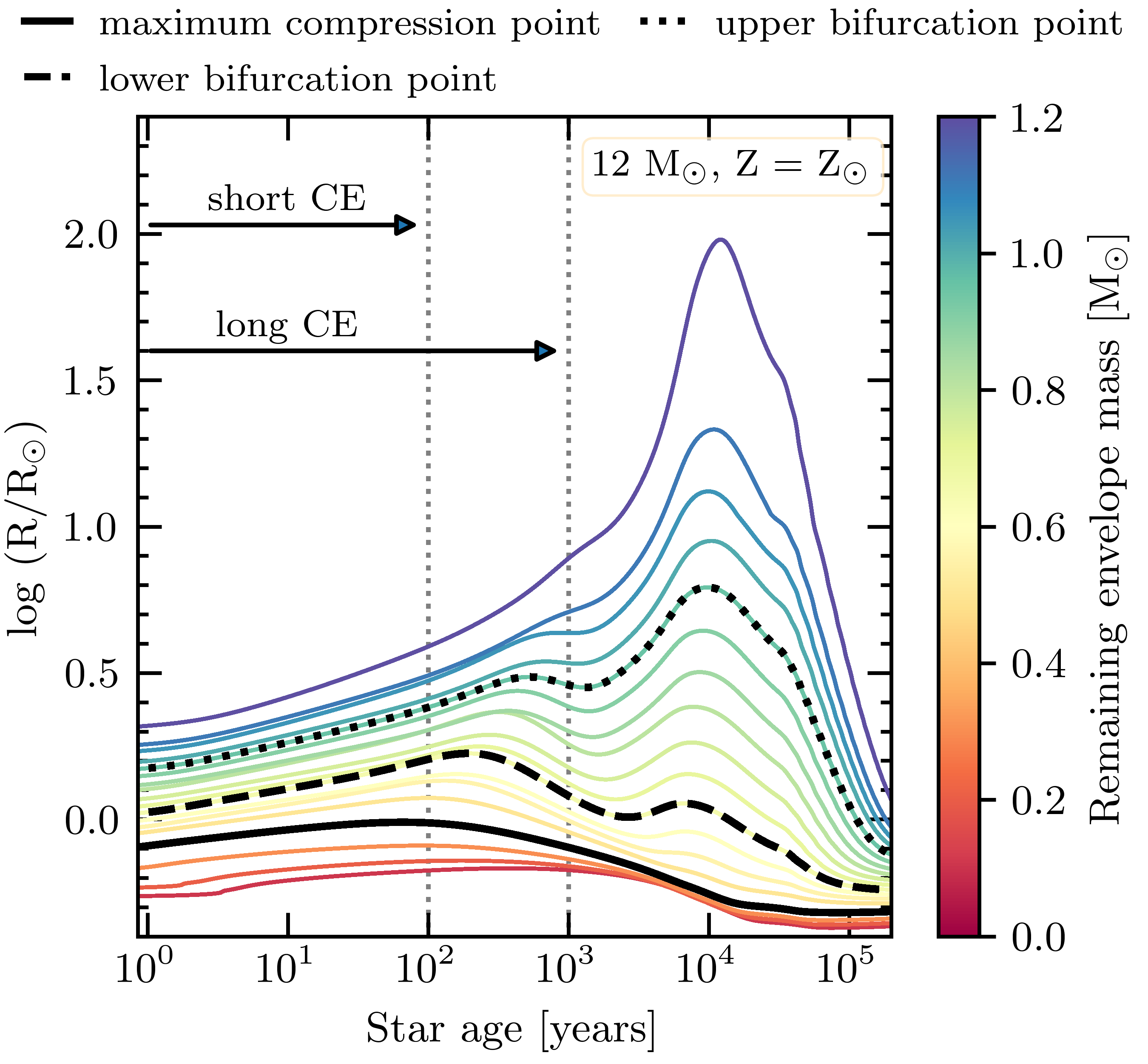}
\caption{
Radial evolution of a stripped (post-\ac{CE} phase) remnant with a 12 $\rm M_{\odot}$ progenitor at $ Z = Z_{\odot}$ (Sec.~\ref{subsub:12}). 
The solid black line indicates the radial evolution of a star stripped to the maximum compression point  ($\approx 0.3\ \rm M_{\odot}$ above the helium core shown in Fig.~\ref{fig:StarProfile12Zsun}). The dashed and dotted black lines indicate the radial tracks for the lower and upper bifurcation points with remaining envelope mass of $\approx 0.65\ \rm M_{\odot}$ and $\approx 0.95\ \rm M_{\odot}$, respectively. 
The black arrows indicate the duration of the short ($t_{100}$) and long ($t_{1000}$) CE phase.
}
\label{fig:Radius12MZS}
\end{figure}
\begin{figure}
\includegraphics[width=1.0\columnwidth]{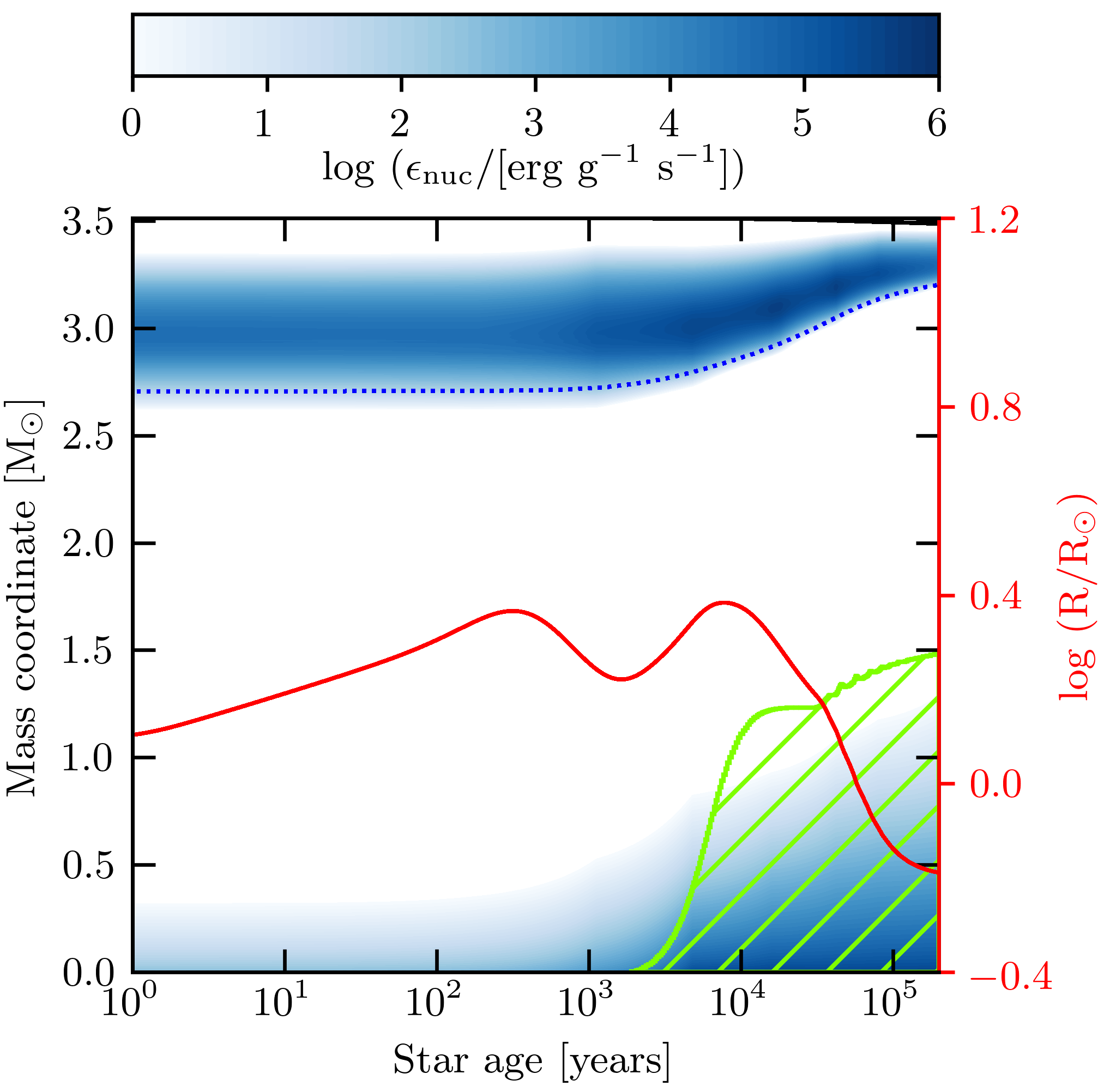}
\caption{
Kippenhahn diagram of a stripped (post-\ac{CE} phase) remnant with a 12 $\rm M_{\odot}$ progenitor at $ Z = Z_{\odot}$ (Sec.~\ref{subsub:12}). 
The stripped remnant has $0.8\ \rm M_{\odot}$ of remaining envelope above the core, in between the lower and upper bifurcation points. 
The plot shows the structure of the star in mass coordinates (left y-axis) from the centre (bottom) to the surface (top) of the star, as a function of time.
Convection is shown as green hatched regions, the dotted blue curve indicates a hydrogen fraction of $X=0.01$, and the intensity of blue indicates the net energy-production rate. 
The radial evolution (right y-axis) is shown as a solid red line. 
}  
\label{fig:kip12}
\end{figure}

\subsection{Post-stripping radial evolution}
\label{sub:radius}
We present the long term evolution of stripped remnants originating from two illustrative cases of massive \ac{CE} donors:
a $12\ \rm{M_{\odot}}$ progenitor at $Z=Z_{\rm{\odot}}$ (Sec.~\ref{subsub:12}) and a $20\ \rm{M_{\odot}}$ progenitor at $Z=Z_{\rm{LMC}}$ (Sec.~\ref{subsub:20}).
The former is stripped at a stage where it has developed a fully convective hydrogen-rich envelope, while the latter has a partially radiative hydrogen-rich  envelope.

\subsubsection{$12\ \rm{M_{\odot}}$ progenitor at $Z=Z_{\odot}$}
\label{subsub:12}
In Fig.~\ref{fig:Radius12MZS} we plot the radial evolution of \ac{CE} remnants for different envelope masses. The lowest remaining envelope masses, below the maximum compression point ($\leq 0.3\ \rm{M_{\odot}}$) lead to negligible expansion on timescales shorter than 1000 yr, and contraction on longer timescales. With increasing remaining envelope mass, two clearly distinguishable expansion phases emerge: a first pulse on the medium (thermal) timescale and a second pulse on a longer timescale.  The first pulse is a result of thermal relaxation of the stripped remnant in response to the partial envelope loss; this part of the evolution is likely to happen when the core--remnant binary is still partially embedded in the \ac{CE}. 

The second maximum in radius is associated with the contraction (expansion) of the helium core (remaining envelope) in the transition to the core-helium burning phase. The second pulse occurs when the luminosity of the remaining hydrogen shell burning reaches a maximum (Fig.~\ref{fig:kip12}). For increasing remaining envelope masses the second expansion becomes more and more significant. For the largest remaining envelope masses, a continuous increase in radius is found and these masses are discarded as possible post-\ac{CE} remnants. All remnants begin to contract at the onset of core helium burning (Fig.~\ref{fig:kip12}), due to a reduction in the amount of remaining hydrogen in the shell.

Assuming the short ($t_{100}$) \ac{CE} criteria, we find that the location of the lower bifurcation point is around remaining envelope masses of $\approx 0.65\ \rm{M_{\odot}}$ and that the radius expands to $\approx 1.5\,\rm{R_\odot}$. More massive envelopes with mass below $\sim 1\,\rm{M_\odot}$ keep expanding, but contract after less than 1000 years. Using the long ($t_{1000}$) \ac{CE} criteria, we find that the upper bifurcation point is around remaining envelope masses of $\approx 0.95\ \rm M_{\odot}$, with a radius just above $2\,\rm{R_\odot}$.

\begin{figure}
\includegraphics[width=1.0\columnwidth]{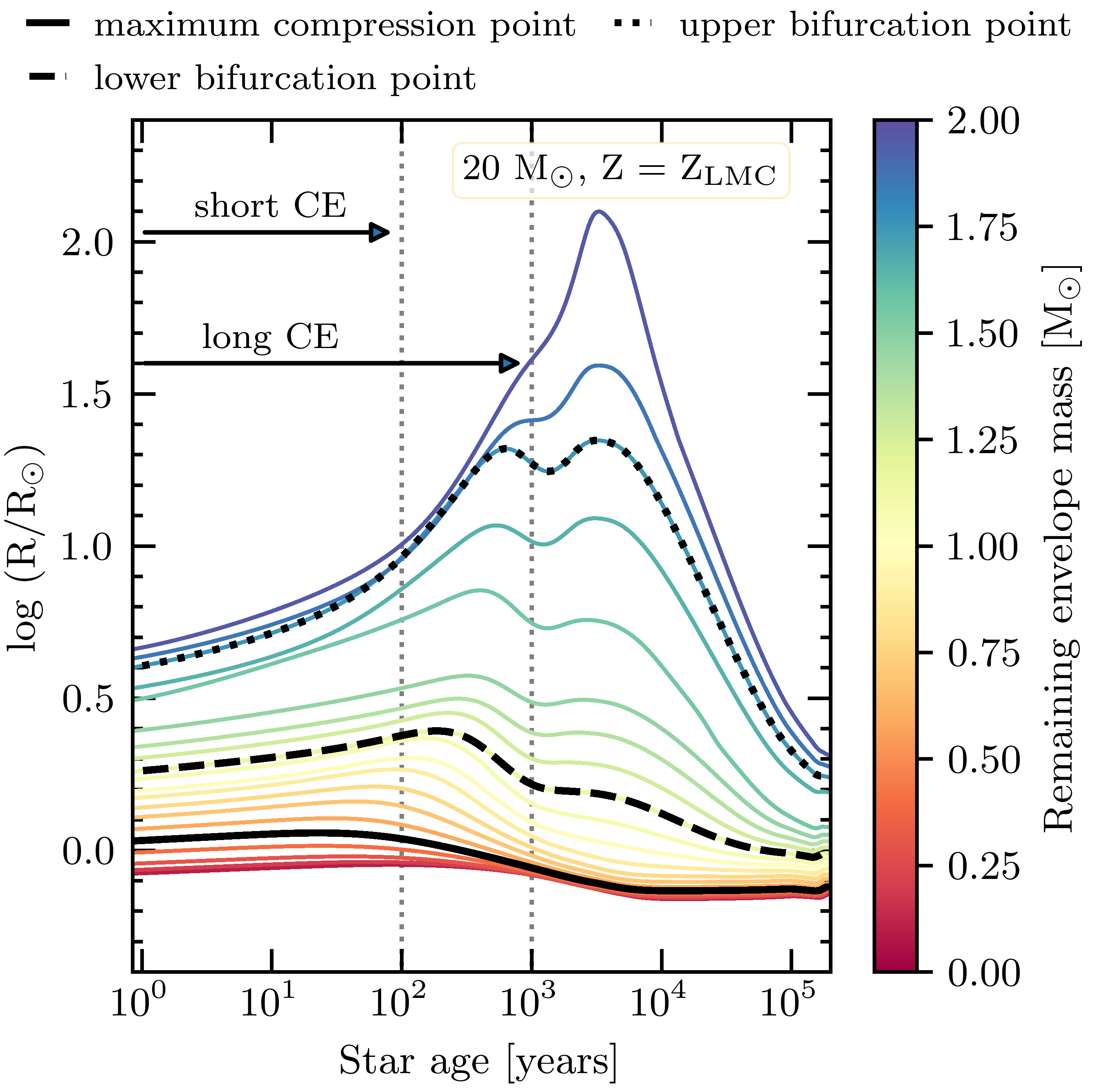}
\caption{
Radial evolution of a stripped (post-\ac{CE} phase) remnant with a 20 $\rm M_{\odot}$ progenitor at $ Z = Z_{\rm{LMC}}$ (Sec.~\ref{subsub:20}).
The labels and descriptions match those from Fig.~\ref{fig:Radius12MZS}.
In this case, the mass coordinate of the maximum compression point is $\approx 0.7\ \rm M_{\odot}$ above the core.
The lower and upper bifurcation points have remaining envelope masses of $\approx 1.2\ \rm M_{\odot}$ and $\approx 1.8\ \rm M_{\odot}$, respectively. 
}
\label{fig:Radius20MZS}
\end{figure}
\begin{figure}
\includegraphics[width=1.0\columnwidth]{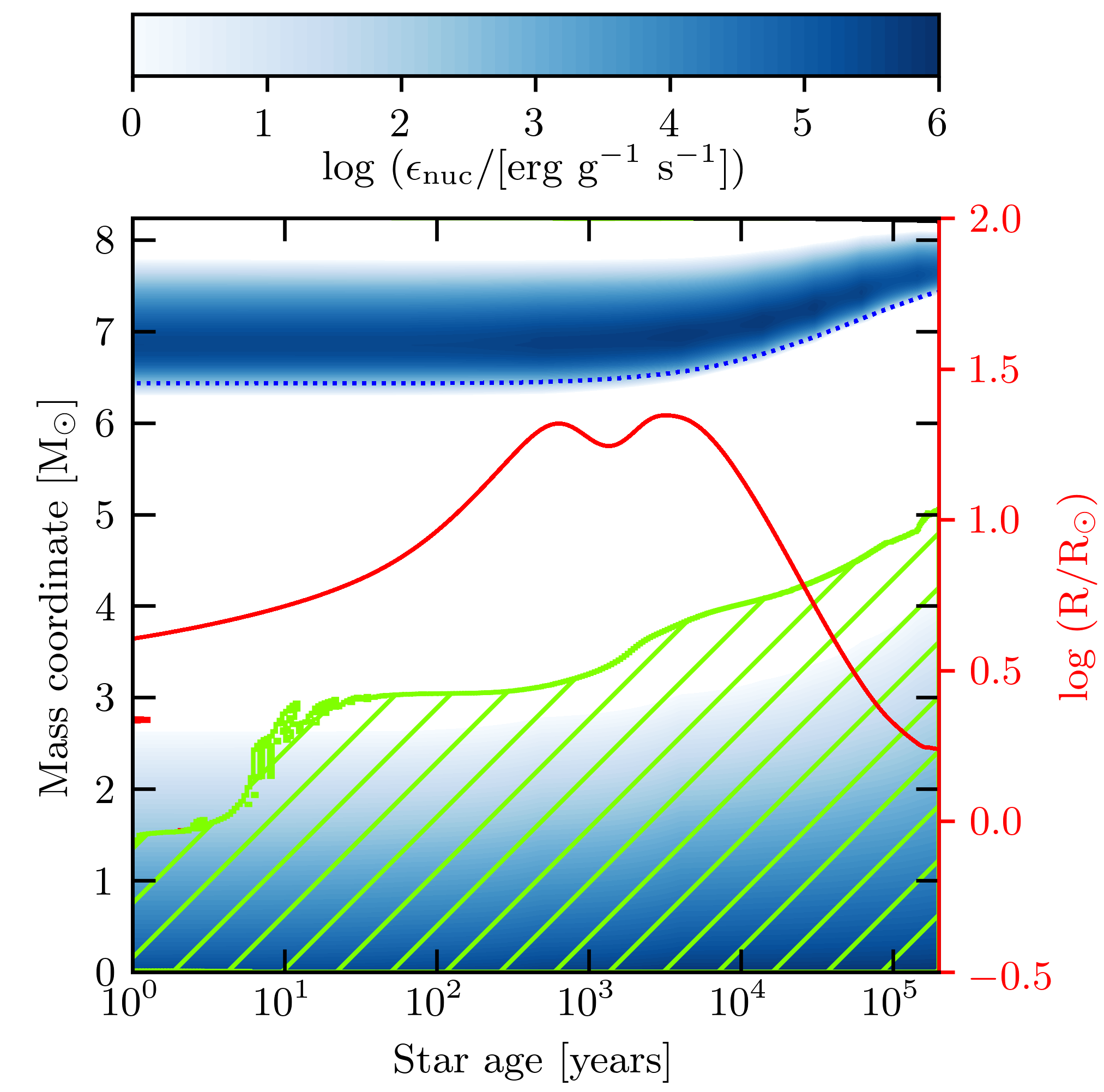}
\caption{
Kippenhahn diagram of a stripped (post-\ac{CE} phase) remnant with a 20 $\rm M_{\odot}$ progenitor at $ Z = Z_{\rm{LMC}}$ (Sec.~\ref{subsub:20}). 
The stripped remnant has $1.8\ \rm M_{\odot}$ of remaining envelope above the core. 
The labels and description match those from Fig.~\ref{fig:kip12}.
}  
\label{fig:kip20}
\end{figure}

\subsubsection{$20\ \rm{M_{\odot}}$ progenitor at $Z=Z_{\rm{LMC}}$}
\label{subsub:20}
In Fig.~\ref{fig:Radius20MZS} we show the radial evolution of the \ac{CE} remnant for different retained envelope masses for a more massive (20~M$_{\odot}$) progenitor at a lower (LMC) metallicity. This stellar model is already fusing helium in the core at the onset of the \ac{CE} phase but the core is not yet in thermal equilibrium and continues to contract (Fig.~\ref{fig:kip20}). 
We see a similar result to the 12~M$_\odot$ star: two different expansion phases, with the second expansion becoming more prominent for more massive retained envelopes, and continued expansion for the most massive envelopes. The second pulse is significantly less pronounced for this stellar mode than for the lower-mass higher-metallicity donor, as long as the mass of the retained envelope is $\lesssim 2\ \rm{M_{\odot}}$. After less than $10^4$ yr all models rapidly contract as the hydrogen-burning shell is depleted.
Stripping below the maximum compression point ($\leq 0.7\ \rm{M_{\odot}}$) leads to negligible expansion on short and medium timescales, and contraction on a longer timescale.
Stripping to the lower bifurcation point, which lies at a remaining envelope mass of  $\approx 1.2\ \rm M_{\odot}$ and radius of $\approx 2\, \rm{R_\odot}$, results in maximum radial expansion reached in the first pulse, followed by an overall contracting behaviour barely perturbed by the second pulse.  Stripping at the upper bifurcation point, with a remaining envelope mass of $\approx 1.8\ \rm M_{\odot}$ and a radius of $\gtrsim 10\ \rm R_{\odot}$, leads to a more noticeable second pulse.

\subsection{Critical points as a function of progenitor mass and metallicity}
\label{sub:comparison}
\begin{figure}
\includegraphics[width=0.48\textwidth]{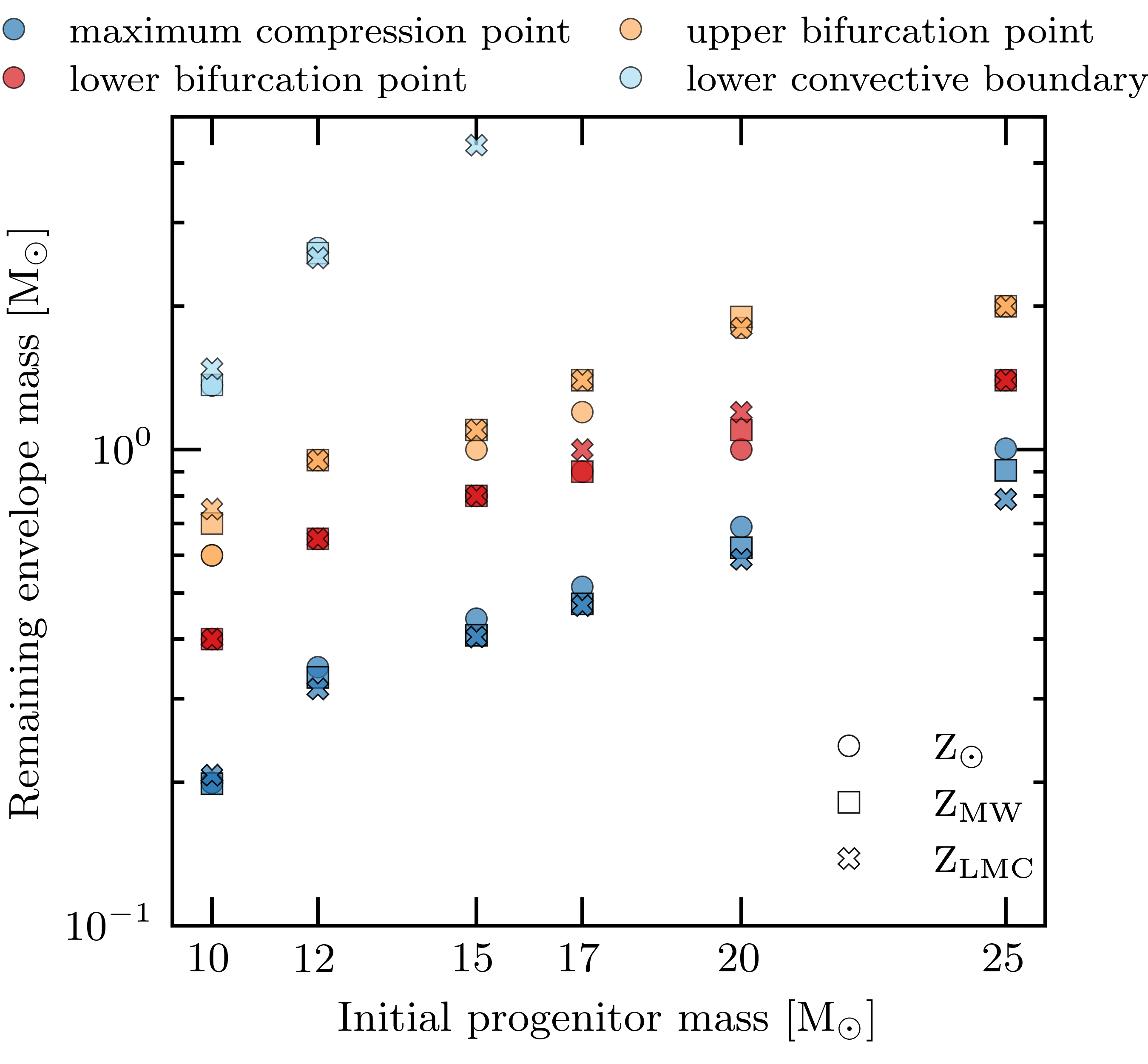}
\caption{
Critical points as a function of initial progenitor mass and remaining envelope mass.
The maximum compression points are shown in navy, the lower bifurcation points are shown in red, the upper bifurcation points are shown in yellow, and the lower boundaries of the convective envelope (if any) are shown in light blue.
Solar metallicity is shown as circles, MW metallicity is shown as squares, and LMC metallicity is shown as crosses.
Tables \ref{tab:lambdaSolar}, \ref{tab:lambdaMW}, and \ref{tab:lambdaLMC} show the value of the core mass for each progenitor mass at each metallicity in this study.}
\label{fig:CritZSZLMC}
\end{figure}

For each of our models, we calculate the maximum compression point and the two bifurcation points as a function of progenitor mass, remaining envelope mass, and metallicity.
We show the results in Fig.~\ref{fig:CritZSZLMC}, which also shows the lower boundary of the outer convective envelope (if present) of the models prior to stripping.

The trend with the initial progenitor mass, the most relevant quantity \textit{a posteriori}, is an overall monotonic increase in the location of the critical points.
Most bifurcation points can be found within the $1\ \rm{M_{\odot}}$ mass shell above the core.
Furthermore, all bifurcation points lie within the $2\ \rm{M_{\odot}}$ mass shell above the core. 
For all models, the bifurcation points have significantly higher envelope masses than the maximum compression points, by factors of $\sim 2$ ($\sim 3$) for the lower (upper) bifurcation point.  This implies that 1 or even 2 M$_\odot$ of hydrogen-rich envelope may be retained by stripped post-\ac{CE} donors.

The effect of metallicity in the calculation of the critical points is subtle.
For the most extreme cases, e.g., the (red) lower bifurcation points of the 17 and 20 $\rm{M_{\odot}}$ models, the contrast between the highest ($Z=Z_{\odot}$) and the lowest ($Z=Z_{\rm{LMC}}$) metallicity results in a difference of $\approx 0.2$ dex in remaining envelope mass.
The least extreme cases, e.g., the upper and lower bifurcation points for the 12 $\rm{M_{\odot}}$ models, result in effectively identical values for all metallicities.
Overall, and in contrast to the variations in mass, the effect of metallicity is not dominant, it does not follows a clear trend, and the estimated values can be considered within the numerical uncertainties. 

\begin{figure}
\includegraphics[width=1.0\columnwidth]{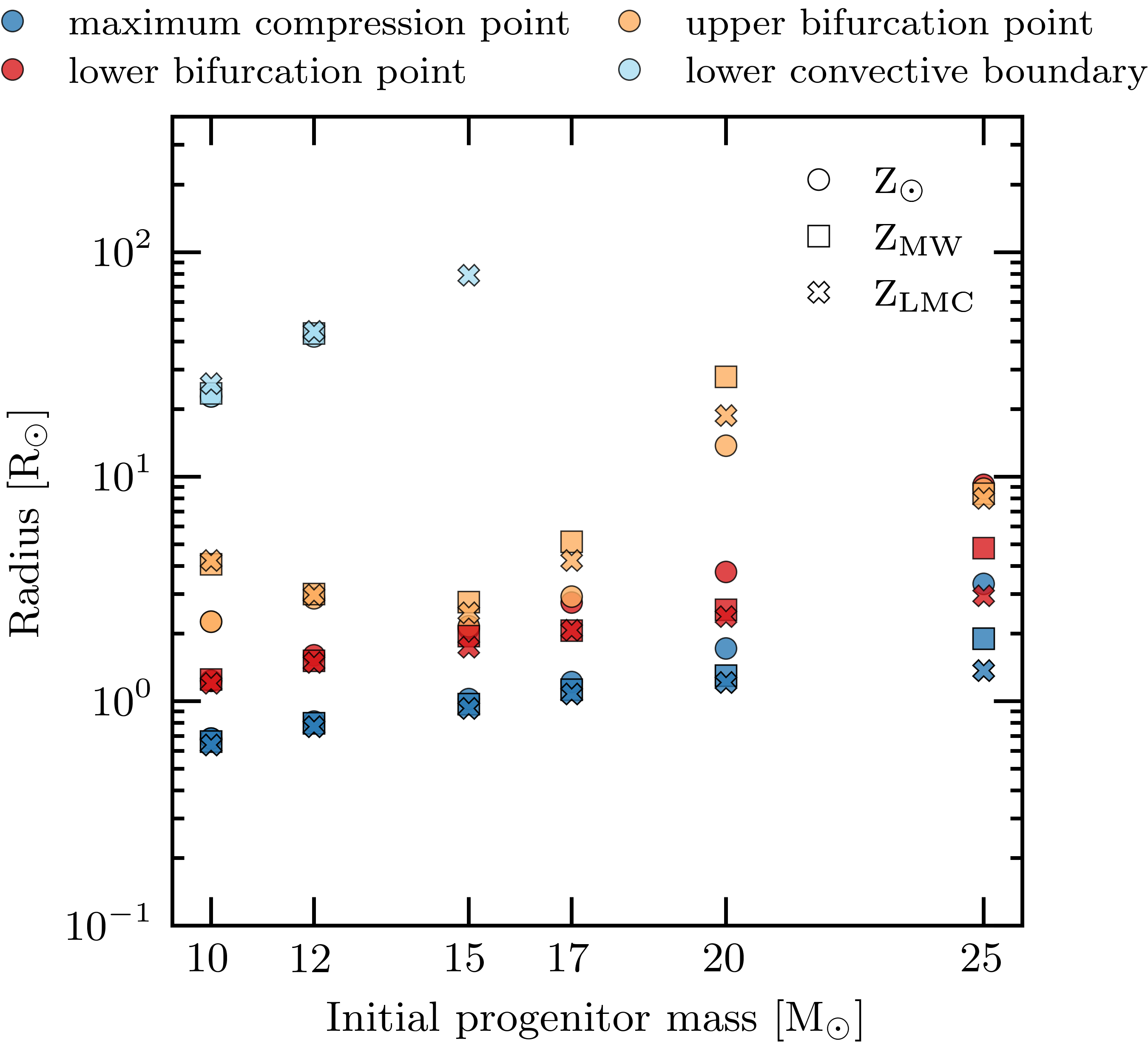}
\caption{
Maximum post-\ac{CE} ($t < 2000$ yr) radial expansion of the stripped models as a function of progenitor mass and metallicity.
The labels and descriptions match those from Fig.~\ref{fig:CritZSZLMC}.
}
\label{fig:CritZSLMCRadius}
\end{figure}

In Fig.~\ref{fig:CritZSLMCRadius}, we show the maximum post-\ac{CE} ($t<2000$ yr) radius as a function of the critical points and progenitor mass. 
Differences of a fraction of a solar mass in remaining envelope masses can lead to variations in radii of several solar radii, and up to an order of magnitude in some cases. This is still smaller than the two orders of magnitude uncertainty in the radii of stripped post-\ac{CE} stars proposed by \citet{KkruckowCEejection}.
Overall, the general trend is monotonically increasing radii as a function of initial progenitor mass, though it is not as clear as the relationship between progenitor and remaining envelope mass.
Stripping to the bottom of the convective envelope always yields remnants with radii larger than $10\ \rm{R_{\odot}}$ for $t<2000$ yr.

\subsection{Estimates of \texorpdfstring{$\lambda$}{1} and \texorpdfstring{$f_R$}{1}: bifurcation points and plausible CE ejection}
\label{sub:alphalambdaresult} 
\begin{figure}
\includegraphics[width=1.0\columnwidth]{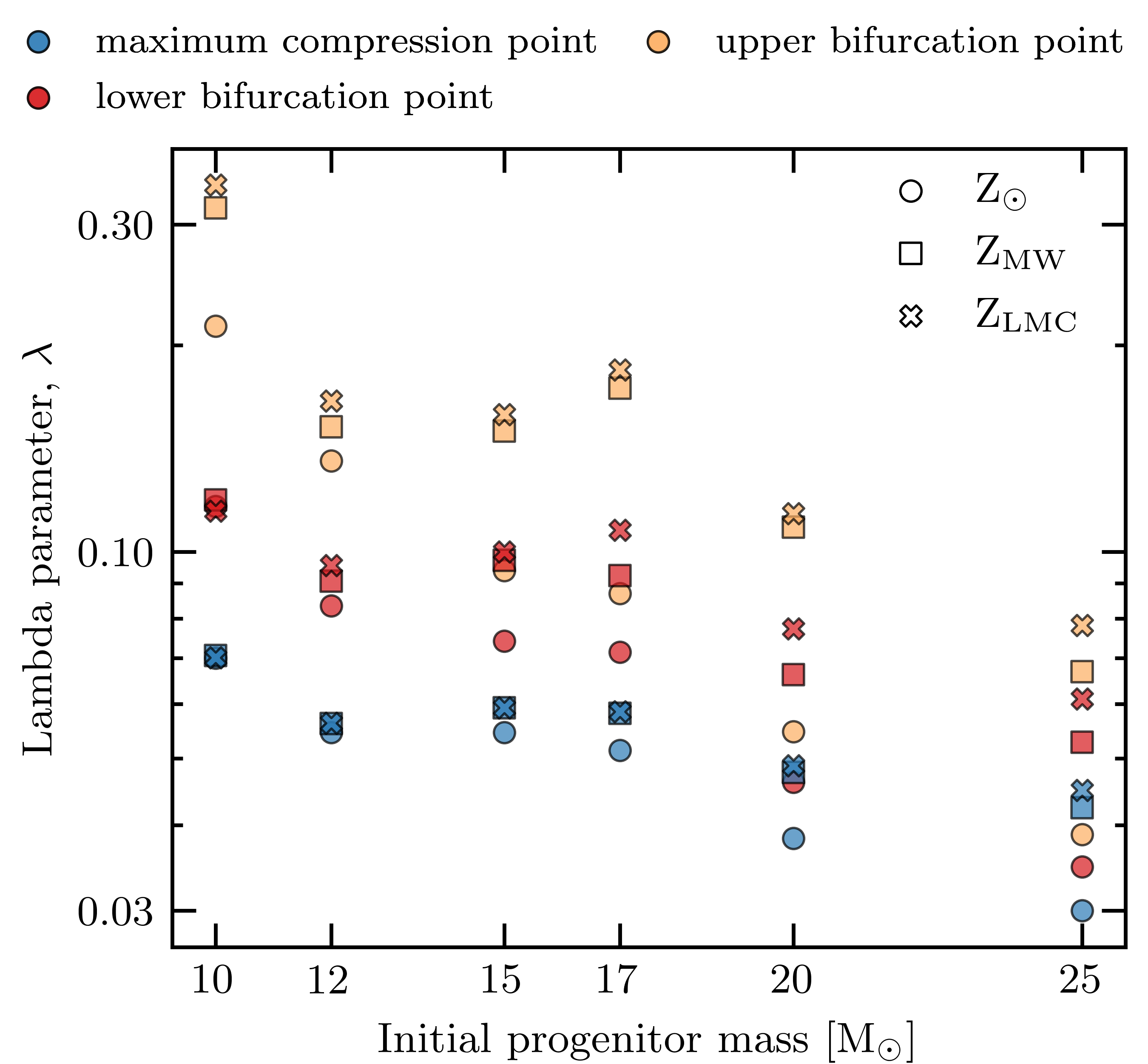}
\caption{
Estimates of the binding energy $\lambda$ parameter at the onset of the \ac{CE} phase depending on the stripping point and progenitor metallicity.
The maximum compression points are shown in navy, the lower bifurcation points in red, and the upper bifurcation point in yellow.
Solar metallicity is shown as circles, MW metallicity is shown as squares, and LMC metallicity is shown as crosses.
}
\label{fig:trendlambda}
\end{figure}

We present $\lambda$ and $f_R$ values, as defined in Sec. \ref{sub:lambdaCalculation}, in Table~\ref{tab:lambdaSolar} for $Z=Z_{\odot}$, Table~\ref{tab:lambdaMW} for $Z=Z_{\rm{MW}}$, and Table~\ref{tab:lambdaLMC} for $Z=Z_{\rm{LMC}}$.
The estimated values of $\lambda$ fall in the range $0.03 \lesssim \lambda \lesssim 0.34$ (Fig.~\ref{fig:trendlambda}), and $f_R>1.0$ for all models.

\begin{table}
\begin{center}
\begin{tabular}{ l l l | c | c | c }
 \hline
 \textbf{Progenitor} & \textbf{Core} & \textbf{Critical} & $\bf M_{\rm{env,rem}}$ & \textbf{$f_R$} & \textbf{$\lambda$} \\
 \textbf{mass} & \textbf{mass} &  \textbf{point} & & \\ 
 \hline
 \hline
 10 & 2.10 & $M_{\rm{cp}}$ & 0.20 & 2.12 & 0.07 \\
 & & $X_H=0.1$ & 0.34 & 1.98 & 0.10 \\  
 & & lower & 0.40 & 2.09 & 0.12 \\
 & & upper & 0.60 & 1.93 & 0.21 \\
 \hline
 12 & 2.71 & $M_{\rm{cp}}$ & 0.35 & 3.15 & 0.05 \\
 & & $X_H=0.1$ & 0.55 & 3.73 & 0.07\\  
 & & lower & 0.65 & 3.72 & 0.08 \\
 & & upper & 0.95 & 3.73 & 0.14 \\
 \hline
 15 & 3.93 & $M_{\rm{cp}}$ & 0.44 & 3.83 & 0.05 \\
 & & $X_H=0.1$ & 0.71 & 4.36 & 0.07 \\  
 & & lower & 0.80 & 4.45 & 0.07 \\
 & & upper & 1.00 & 3.43 & 0.09 \\
 \hline
 17 & 4.81 & $M_{\rm{cp}}$ & 0.52 & 4.03 & 0.05 \\
 & & $X_H=0.1$ & 0.82 & 5.28 & 0.07\\  
 & & lower & 0.90 & 5.52 & 0.07 \\
 & & upper & 1.20 & 4.46 & 0.09 \\
 \hline
 20 & 6.12 & $M_{\rm{cp}}$ & 0.69 & 6.26 & 0.04 \\
 & & $X_H=0.1$ & 1.00 & 7.26 & 0.05\\  
 & & lower & 1.00 & 8.35 & 0.05 \\
 & & upper & 1.80 & 8.49 & 0.05 \\
 \hline
 25 & 8.36 & $M_{\rm{cp}}$ & 1.00 & 7.67 & 0.03 \\
 & & $X_H=0.1$ & 1.45 & 12.18 & 0.03\\  
 & & lower & 1.40 & 14.35 & 0.03 \\
 & & upper & 2.00 & 13.19 & 0.04 \\
 \hline
\end{tabular}
\end{center}
\caption{
Estimates of the $\alpha$ and $\lambda$ parameters, for all $Z=Z_{\odot}$ models, assuming different critical points for the core/envelope boundary: the maximum compression point ($M_{\rm{cp}}$), the lower bifurcation point (lower), and the upper bifurcation point (upper).
$M_{\rm{env,rem}}$ corresponds to the remaining envelope mass for each critical point.
The core mass corresponds to $X_H\approx10^{-6}$.
All masses are in solar units.
}
\label{tab:lambdaSolar}
\end{table}

\begin{table}
\begin{center}
\begin{tabular}{ l l l | c | c | c }
 \hline
 \textbf{Progenitor} & \textbf{Core} & \textbf{Critical} & $\bf M_{\rm{env,rem}}$ & \textbf{$f_R$} & \textbf{$\lambda$} \\
 \textbf{mass} & \textbf{mass} &  \textbf{point} & & \\ 
 \hline
 \hline
 10 & 2.21 & $M_{\rm{cp}}$ & 0.20 & 1.97 & 0.07 \\
 & & $X_H=0.1$ & 0.30 & 2.05 & 0.09 \\   
 & & lower & 0.40 &  2.01 & 0.12 \\
 & & upper & 0.70 &  2.21 & 0.32 \\
 \hline
 12 & 2.82 & $M_{\rm{cp}}$ & 0.33 & 3.07 & 0.06 \\
 & & $X_H=0.1$ & 0.51 & 3.36 & 0.07 \\   
 & & lower & 0.65 &  3.19 & 0.09 \\
 & & upper & 0.95 &  3.45 & 0.15 \\
 \hline
 15 & 4.07 & $M_{\rm{cp}}$ & 0.41 &  3.40 & 0.06 \\
 & & $X_H=0.1$ & 0.65 & 3.76 & 0.08  \\   
 & & lower & 0.80 &  3.77 & 0.10 \\
 & & upper & 1.10 &  3.25 & 0.15 \\
 \hline
 17 & 4.96 & $M_{\rm{cp}}$ & 0.47 &  3.92 & 0.06 \\
 & & $X_H=0.1$ & 0.74 & 4.42 & 0.08 \\   
 & & lower & 0.90 &  4.18 & 0.09 \\
 & & upper & 1.40 &  5.13 & 0.17 \\
 \hline
 20 & 6.33 & $M_{\rm{cp}}$ & 0.62 &  5.45 & 0.05 \\
 & & $X_H=0.1$ & 0.92 & 6.77 & 0.06 \\   
 & & lower & 1.10 &  7.17 & 0.07 \\
 & & upper & 1.90 &  48.75 & 0.11 \\
 \hline
 25 & 8.69 & $M_{\rm{cp}}$ & 0.91 &  8.64 & 0.04 \\
 & & $X_H=0.1$ & 1.25 & 12.09 &  0.05 \\   
 & & lower & 1.40 &  16.78 & 0.05\\
 & & upper & 2.00 &  22.02 & 0.07\\
 \hline
\end{tabular}
\end{center}
\caption{
Same as Table \ref{tab:lambdaSolar} but for $Z=Z_{\rm{MW}}$.}
\label{tab:lambdaMW}
\end{table}

\begin{table}
\begin{center}
\begin{tabular}{ l l l | c | c | c }
 \hline
 \textbf{Progenitor} & \textbf{Core} & \textbf{Critical} & $\bf M_{\rm{env,rem}}$ & \textbf{$f_R$} & \textbf{$\lambda$} \\
 \textbf{mass} & \textbf{mass} &  \textbf{point} & & \\ 
 \hline
 \hline
 10 & 2.30 & $M_{\rm{cp}}$ & 0.21 & 1.94 & 0.07 \\
 & & $X_H=0.1$ & 0.28 & 2.25 & 0.08 \\    
 & & lower & 0.40 & 2.05 & 0.11 \\
 & & upper & 0.75 & 2.14 & 0.34 \\
 \hline
 12 & 2.90 & $M_{\rm{cp}}$ & 0.31 &  2.91 & 0.06 \\
 & & $X_H=0.1$ & 0.49 & 3.28 & 0.07 \\    
 & & lower & 0.65 & 2.94 & 0.10 \\
 & & upper & 0.95 & 3.11 & 0.17 \\
 \hline
 15 & 4.15 & $M_{\rm{cp}}$ & 0.40 &  3.26 & 0.06 \\
 & & $X_H=0.1$ & 0.62 & 3.40 & 0.08 \\    
 & & lower & 0.80 & 3.29 & 0.10 \\
 & & upper & 1.10 & 2.78 & 0.16 \\
 \hline
 17 & 5.05 & $M_{\rm{cp}}$ & 0.47 & 3.78 & 0.06 \\
 & & $X_H=0.1$ & 0.69 & 3.90 & 0.07\\    
 & & lower & 1.00 & 3.60 & 0.11 \\
 & & upper & 1.40 & 4.02 & 0.18 \\
 \hline
 20 & 6.43 & $M_{\rm{cp}}$ & 0.59 & 5.09 & 0.05 \\
 & & $X_H=0.1$ & 0.85 & 6.55 & 0.06\\    
 & & lower & 1.20 & 5.77 & 0.08 \\
 & & upper & 1.80 & 30.95 & 0.11 \\
 \hline
 25 & 8.87 & $M_{\rm{cp}}$ & 0.79 & 6.29 & 0.04 \\
 & & $X_H=0.1$ & 1.15 & 8.27 & 0.05 \\    
 & & lower & 1.40 & 9.27 & 0.06 \\
 & & upper & 2.00 & 18.77 & 0.08 \\
 \hline
\end{tabular}
\end{center}
\caption{
Same as Table \ref{tab:lambdaSolar} but for $Z=Z_{\rm{LMC}}$.}
\label{tab:lambdaLMC}
\end{table}

\begin{figure}
\includegraphics[width=1.0\columnwidth]{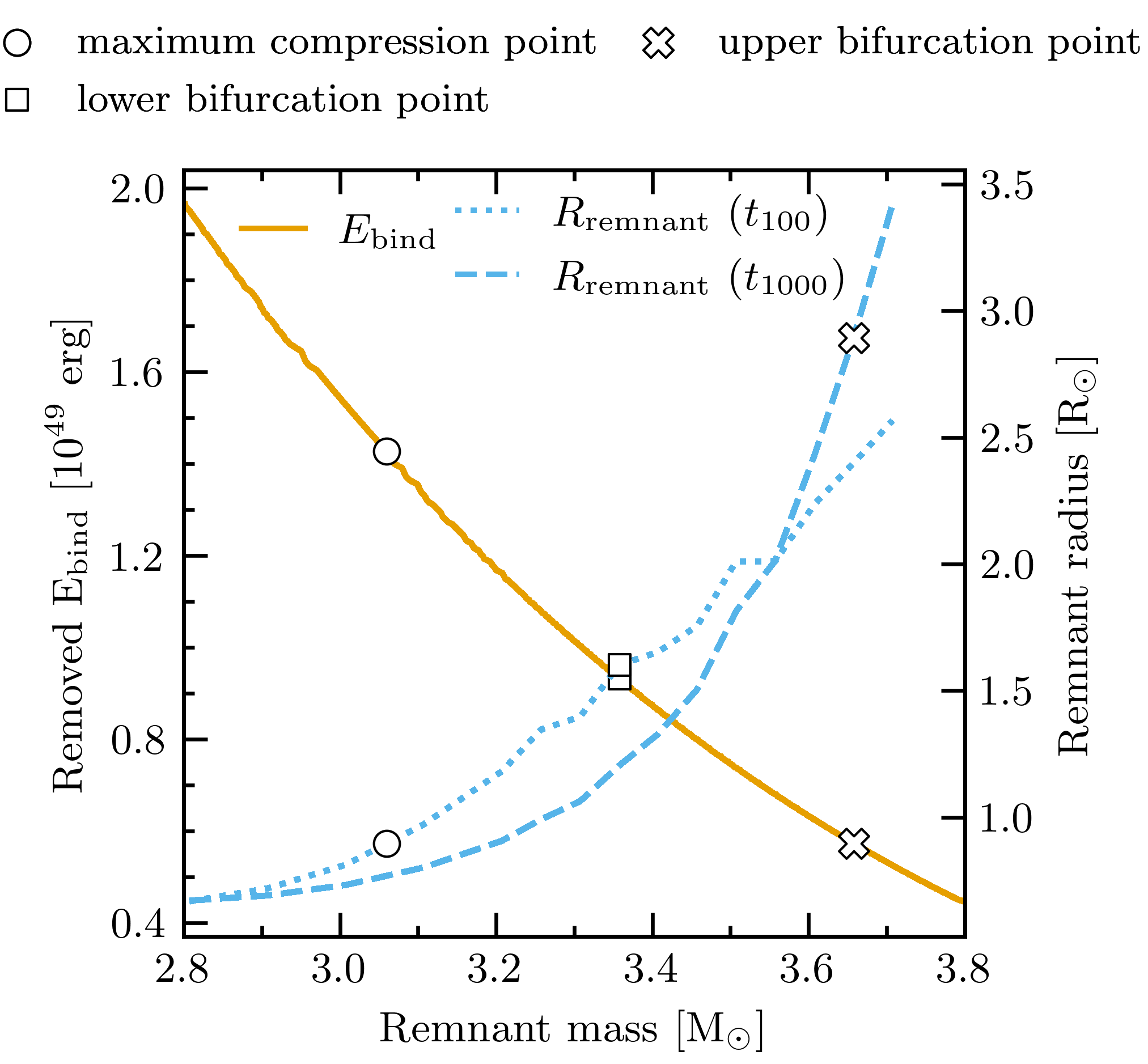}
\caption{
The initial binding energy of the ejected envelope (solid ochre) and radius of the stripped remnant at $t_{100}$ (dotted cyan) or $t_{1000}$ (dashed cyan), plotted as a function of different remnant masses from a $12\ \rm{M_{\odot}}$ progenitor at $Z = Z_{\odot}$. 
Symbols denote the maximum compression point (circle), as well as the lower (square) and upper (cross) bifurcation points.
}
\label{fig:trendradiusbinding}
\end{figure}

Tables~\ref{tab:lambdaSolar}, \ref{tab:lambdaMW}, and \ref{tab:lambdaLMC} reveal that the larger the remaining envelope mass the larger the $\lambda$ value, i.e. the lower the binding energy of the ejected envelope. This makes intuitive sense and illustrates the sensitivity of the envelope binding energy to the bifurcation point location, as discussed in the literature (Sec.~\ref{sec:Introduction}). However, in the majority of cases, the increase in the $\lambda$ values does not necessarily makes the \ac{CE} ejection more feasible. 
This is because an increase of the remnant mass, apart from lowering the envelope binding energy, also leads to an increase of the remnant radius. We illustrate this in Fig.~\ref{fig:trendradiusbinding}, using the 12 $\rm{M_{\odot}}$ donor at $Z=Z_{\rm{MW}}$ as an example. 
The larger the remnant, the larger the post-\ac{CE} binary separation needs to be to form a detached binary.

\begin{figure}
\includegraphics[trim=12.0 0 0 0, clip,width=1.0\columnwidth]{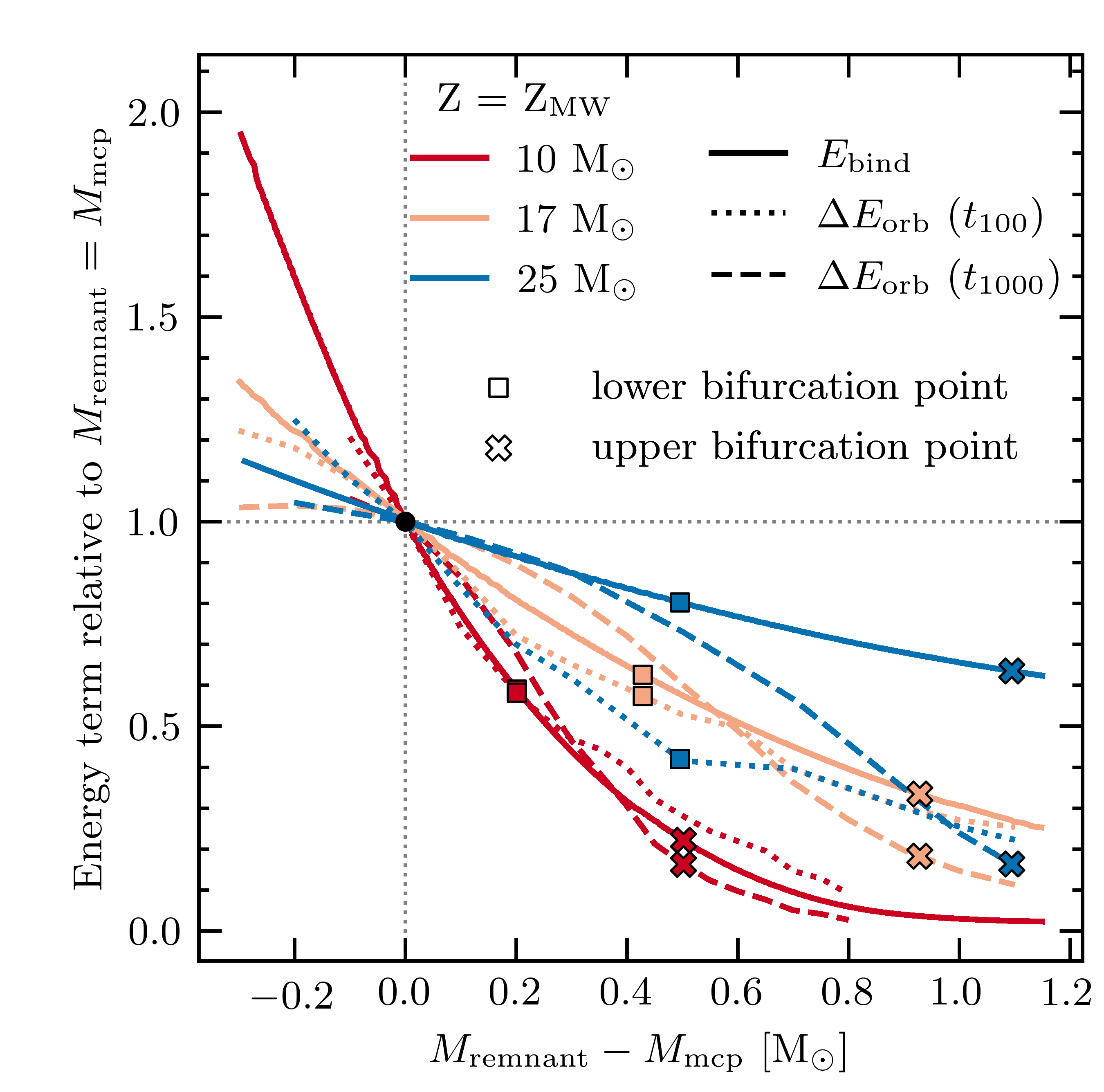}
\caption{
The binding energy (solid curves) and the change in orbital energy (dotted and dashed curves) normalised to their values at the maximum compression point.  The orbital energy change is computed by assuming that the final separation is determined by $R_{\rm{RL,f}} \, =\, R_\mathrm{stripped}$ at $t_{100}$ (dotted) or $t_{1000}$ (dashed) after stripping.
The mass coordinate of the 10, 17, and 25 M$_\odot$ stars at $Z=Z_\mathrm{MW}$ is also given relative to the maximum compression point.
The condition for successful ejection is $\alpha \times \Delta E_{\textrm{orb}} > E_{\textrm{bind}}$.
}
\label{fig:trendenergies}
\end{figure}

We briefly explore the effect of the energy formalism in determining the post-\ac{CE} separation of the binary.
For an increasing remnant mass, the decrease in the energy source term ($\Delta E_{\rm orb}$) will counteract the decrease in the energy sink ($E_{\rm bind}$). We illustrate this in Fig.~\ref{fig:trendenergies}, where we plot the relative change in the binding energy (solid curves) and the energy from orbital inspiral (dotted and dashed curves) normalized to their values at the maximum compression point. For the 10 $\rm{M_{\odot}}$ and 17 $\rm{M_{\odot}}$ donor cases, the relative change in $\Delta E_{\rm orb}$ and $E_{\rm bind}$ is similar for a wide range of remnant masses. 

In the case of the most massive donors (as illustrated by the 25 $\rm{M_{\odot}}$ example in Fig.~\ref{fig:trendenergies}), the decrease in the energy source ($\Delta E_{\rm orb}$) is far more substantial than the decrease in the energy sink ($E_{\rm bind}$).
Altogether, Fig.~\ref{fig:trendenergies} shows why the ``gain" from the binding energy decrease does not necessarily lead to more likely \ac{CE} ejection, according to the \ac{CE} energy budget considerations.

\section{Discussion and Conclusions}
\label{sec:discussion}
We discuss the caveats and limitations of our method (Sec.~\ref{sub:caveats}), compare our estimates of $\lambda$ to the literature (Sec.~\ref{sub:alphalambda}), discuss the implications of our findings  (Sec.~\ref{sub:binaries}) and finish with a summary of our conclusions (Sec.~\ref{sub:conclusions}).

\subsection{Caveats and limitations}
\label{sub:caveats}
We investigated the properties of possible post-\ac{CE} remnants, but with our method we could not determine if we really would expect these to be the end states after the \ac{CE}. In particular, we made no attempt to model the \ac{CE} episode itself, including whether sufficient energy is, in fact, available to eject the envelope according to the budget we computed, or the duration of the various phases of the interaction.

The \ac{CE} is intrinsically a three-dimensional phenomenon. Drag, hydrodynamical instabilities, shock waves, accretion, and radiation transport will play a role on the evolution of the \ac{CE} phase, and they can only be simplified in lower dimensions. Our stellar models are calculated in one dimension and non-spherical effects are not considered. Furthermore, we assumed instantaneous stripping and then allow for either 100 or 1000 yr further evolution assuming that in reality this would happen during the \ac{CE}, while the donor is expanding. As most of the binding energy is stored in the inner layers of the envelope \citep[e.g.,][]{PhDDewi, Klencki2021A&A}, the response of these layers and the mechanical work they do on the envelope can significantly overestimate the energy required for envelope ejection. It is thought that this correction is likely insignificant for degenerate cores \citep{IvanovaReview}, but as our heavier stars cores are not degenerate it could be more important. Finally,  variations in the definition of a successful ejection, such as the duration of the post-\ac{CE} phase and expansion after stripping, will lead to different values for the bifurcation points.
In addition to these caveats there are other limitations to our study. Here we list them in no particular order.

\subsubsection{Energy terms in the energy budget considerations
}
The envelope binding energies we calculate  (Eq.~\ref{eqn:bindingEnergy}) consist of the  gravitational binding energy, reduced by the internal energy (including energy from recombination of hydrogen and helium).
In the \ac{CE} energy budget considerations, we equate that to the energy provided by the orbital shrinkage.
Possible additional energy sources are thus ignored: nuclear energy input from a stellar companion \citep{Podsiadlowski2010}, enhanced burning in shells, or accretion energy from matter retained by the compact companion \citep{1993PASP..105.1373I}, and stationary mass outflows from dynamical instabilities in the convective envelope \citep[the \textit{enthalpy consideration} from][]{2011ApJ...731L..36I}.
On the other hand, we also neglect possible energy sinks, in particular radiative losses from the \ac{CE} surface and terminal kinetic energy of the ejecta \citep{IvanovaReview}.

\subsubsection{Single stellar models for the donors}
Throughout this paper we use single stellar models that simplify the binary element of the \ac{CE} phase.
We evolved the donor from the zero-age main sequence to the onset of the \ac{CE} phase as a single star (Sec.~\ref{sub:progenitor}), but this might not be the case for, e.g., \ac{BNS}-forming systems.
In the early phases of most \ac{BNS}-forming binaries \citep[e.g.,][]{BHstandaardscenario,TaurisFormation,VignaGomez2018}, the stellar progenitor of the first-born \ac{NS} donates mass to the main-sequence companion; this companion will eventually be the giant donor in the \ac{CE} phase \citep[e.g.,][]{2020PASA...37...38V}.
Therefore, our model is either simplifying the response of the main sequence accretor in this early mass transfer episode or it corresponds to an alternative formation channel.
Regardless, this is likely not the main source of uncertainty during the \ac{CE} phase.

Our emulation of the \ac{CE} phase itself is also done using a single stellar model (Sec~\ref{sub:stripping}). 
We do not consider a companion and do not self-consistently model the main phases of the \ac{CE} episode \citep[loss of co-rotation, the plunge-in, and the self-regulated spiral-in, as described by, e.g.,][]{IvanovaCEPitfalls,IvanovaReview}.
We assume the giant is in an orbital configuration that will lead to a \ac{CE} phase and simplify the ablation of the envelope.

Finally, we also do not explicitly consider the presence of the \ac{NS} companion after the \ac{CE} phase. 
The post-\ac{CE} orbit could result in additional mass transfer episodes \citep[e.g.,][]{TaurisFormation,VignaGomez2018} or alternative formation scenarios \citep[e.g.,][]{Vigna-Gomez:2021b}.

\subsubsection{Single donor radii at the onset of the \ac{CE} phase}
While we explored a range of progenitor masses and metallicities, we only considered stars with a radius of 500 $\rm{R_{\odot}}$ at the onset of Roche lobe overflow leading to the \ac{CE}.
In reality, the donor radius at Roche lobe overflow will depend on the separation, the eccentricity, and the mass of the \ac{NS} companion, and possibly stellar rotation, which we ignored here.
This results in a wide variety of massive binaries engaging in a \ac{CE} phase: most of them are giants \citep{VignaGomez2018,2020PASA...37...38V,Klencki2020A&A} and some of them even have residual eccentricity at the moment the \ac{CE} phase begins \citep{2020PASA...37...38V,2021MNRAS.503.5569V}.
Importantly, at the chosen size (500 $\rm{R_{\odot}}$) the lower-mass ($\leq 15\ \rm{M_{\odot}}$) progenitors have a deep outer convective envelope, while the higher-mass ($\geq 17\ \rm{M_{\odot}}$) progenitors, still crossing the Hertzsprung gap, have an outer radiative envelope.
Massive stars that engage in a \ac{CE} phase and have a deep convective envelope are more likely to lead to a successful envelope ejection in contrast to those with a radiative envelope \citep{Klencki2021A&A}.
Our method self-consistently estimates the maximum compression and bifurcation points of our models in order to make a direct comparison between them; however, we reinforce that the method is not a comprehensive study of \ac{CE} evolution.

\subsubsection{Entropy of stripped stars}
Our stripping method ignores the deposition of entropy between the onset of the \ac{CE} phase and the moment when the star is fully stripped, and only relaxes the stellar model once the envelope has been removed (Sec. \ref{sub:stripping}). 
However, it is likely that the entropy of stripped stars does not change as abruptly as in our method.
Any deposition of entropy will imply a larger expansion in the first thermal relaxation phase, prior to envelope stripping.
Therefore, our method likely underestimates the radius of the star immediately after stripping aka after the dynamical inspiral phase.
This suggests that the compression points from our models can be deeper in the star.
On the other hand, any residual additional entropy in the outer layers of the stripped remnant with respect to our method would likely lead to a more pronounced expansion of the remnant during the thermal relaxation following the phase of dynamical inspiral. 
This could be tested in the future by including artificial entropy injection prior to the relaxation phase.

\subsection{Comparison of our \texorpdfstring{$\lambda$}{1} values with the literature}
\label{sub:alphalambda}
We estimate a broad range of $\lambda$ values in this paper for various choices of metallicity, progenitor mass, and core/envelope boundaries.  The overall range across all models is $0.03 \lesssim \lambda \lesssim 0.34$ (Sec.~\ref{sub:alphalambdaresult} and Fig.~\ref{fig:trendlambda}).
When the core/envelope boundaries are placed deeper in the star, the binding energies are greater, and thus the values of $\lambda$ are lower.
Thus, evaluating $\lambda$ at $M_{\rm cp}$ yields $0.03 \lesssim \lambda \lesssim 0.07$; at the lower bifurcation point, $0.03 \lesssim \lambda \lesssim 0.12$; and at the upper bifurcation point, $0.04 \lesssim \lambda \lesssim 0.34$.

Comparisons of $\lambda$ across different studies are challenging because of different assumptions regarding progenitor masses and metallicities, models of stellar evolution, and the evolutionary phases (as parameterised, e.g., by stellar radii) at the time when $\lambda$ are computed.  Perhaps the largest difference is in the core/envelope boundary.  Most studies use definitions based on the hydrogen mass fraction $X_H$, e.g., $X_H=0.1$.  This is often, but not always, close to $M_{\rm cp}$ (see, e.g., Fig.~\ref{fig:StarProfile12Zsun}, and Fig.~5 of \citealt{IvanovaReview}), so we generally compare the values in the literature against our values at $M_{\rm cp}$. 

\citet{DewiTauris2000} computed values of $\lambda$ for stars with masses $3-10\ \rm{M_{\odot}}$ at a radius of $500\ \rm{R_{\odot}}$, assuming that the core/envelope boundary is at $X_H\approx 0.10$.
The most direct comparison we can make is between their 10 $\ \rm{M_{\odot}}$ model and our 10 $\ \rm{M_{\odot}}$ model at $Z=Z_{\odot}=0.017$.
The model of \citet{DewiTauris2000} leads to values of $\lambda=0.35$ or $\lambda=0.17$, depending on whether or not internal energy is included, respectively.
Our model, which includes internal energy, leads to $0.07 \leq \lambda \leq 0.21$, depending on the core/envelope boundary definition. 
Overall, for the 10 $\ \rm{M_{\odot}}$ model, the value of $\lambda$ differs by a factor of a few ($\lesssim 5$).
The discrepancy is likely due to different choices for the metallicity,  different assumptions for stellar evolution parameters (e.g., they use a larger $\alpha_{\textrm{MLT}} = 2.0$), and the definition of the core/envelope boundary.

\cite{2001ASPC..229..255D} extended the models from \cite{DewiTauris2000} to higher masses. 
While these results are not tabulated, one can estimate the values of $\lambda$ at $500\ \rm{R_{\odot}}$ from their plots.
Their 15 $\rm{M_{\odot}}$ model leads to $\lambda \approx 0.1$, closer to the $\lambda$ value from the upper bifurcation point in our study ($\lambda = 0.09$).
The 20 and 25 $\rm{M_{\odot}}$ models lead to $\lambda \approx 0.05$, which is in good agreement with the critical points from both our models, which lead to $0.03 \leq \lambda \leq 0.05$.

\cite{Tauris2001} followed \cite{DewiTauris2000,2001ASPC..229..255D} and studied how different bifurcation points lead to different values of $\lambda$. 
The only model we can somewhat compare with from that study is the 10 $\rm{M_{\odot}}$ model at $Z=0.02$.
The $\lambda$ estimates for that model, which include binding energy, are calculated at the tip of the red giant branch ($374\ \rm{R_{\odot}}$) and the tip of the asymptotic giant branch ($588\ \rm{R_{\odot}}$), leading to values between $0.10 \leq \lambda \leq 3.20$.
Our better agreement is for models where the core/envelope boundary is assumed to be $X_H<0.10$ and closer to the tip of the red giant branch.

\cite{PodsiadBHB} computed the evolution of $\lambda$ as a function of stellar radii and assumed that the core mass is the central mass that contains 1 $\rm{M_{\odot}}$ of hydrogen.
This definition leads to estimated values between $0.04 \lesssim \lambda \lesssim 0.07$ for models with $20$ and $25\ \rm{M_{\odot}}$ donors at 500 $\rm{R_{\odot}}$ (see their Fig.~1, panel 2).
These values are in agreement with our $\lambda$ estimates at the maximum compression point at any metallicity.

\cite{2010ApJ...716..114X,2010ApJ...722.1985X} made an exhaustive study of the binding energy and the $\lambda$ parameter in the context of high- (Pop I) and low- (Pop II) metallicity environments.
They assumed that the core/envelope boundary is at $X=0.15$.
We compare their Pop I and II metallicity values with our $Z=Z_{\odot}$ and $Z=Z_{\rm{LMC}}$ results, respectively.
Our estimates, either considering the maximum compression or the lower bifurcation point as a reference, agree well within a factor of a few.
However, for the $10\ \rm M_{\odot}$ progenitor, they estimate values between $0.3 \leq \lambda \leq 1$, generally larger than any of our estimates of $\lambda \leq 0.34$ for that same model. 

More recently, \cite{KkruckowCEejection} carried out a study of the binding energy in the context of gravitational-wave sources, particularly binary black hole mergers. 
For their $15\ \rm M_{\odot}$ model at Milky Way metallicity, they estimate $\lambda \approx 0.06$ at $500\ \rm{R_{\odot}}$ (see upper panel of their Fig. 1).
This value is in good agreement with our estimate close to the maximum compression point (Table \ref{tab:lambdaMW}).
Their more massive $26\ \rm M_{\odot}$ model at Milky Way metallicity also seems to roughly agree with our $25\ \rm M_{\odot}$ estimate (see upper panel of their Fig. 1).

Also in the context of gravitational-wave sources, \cite{Klencki2021A&A} studied binding energies of massive giants across a wide range of masses and metallicities.
For their $16\ \rm M_{\odot}$ and $25\ \rm M_{\odot}$ models at $500\ \rm{R_{\odot}}$, they find $\lambda$ values in the range $0.04 \leq \lambda \leq 0.08$, depending on metallicity (cf. their Fig. B.3). This is in good agreement with our estimates that assume the core/envelope boundary at maximum compression point. Both \cite{KkruckowCEejection} and \cite{Klencki2021A&A} assumed $X_H = 0.1$ for the bifurcation points and noted that a similar result would be obtained for boundaries at maximum compression points.

Finally, \cite{2021A&A...650A.107M} studied the implications of binding energy in mass transfer of binary black hole progenitor systems.
They consider the detailed evolution of a $30\ \rm M_{\odot}$ model at $Z=Z_{\odot}/10$.
They find that, at $500\ \rm{R_{\odot}}$, different parameterisation and fitting formulae lead to uncertainties in binding energies between $48 \lesssim \log_{10} (-E_{\rm{bind}}/\rm{erg}) \lesssim 50$ (cf. their Fig.~6), which propagate directly to uncertainties of two orders of magnitude in $\lambda$.

\cite{2010ApJ...717..724G} proposed an alternative definition of the binding energy.  They suggested that in order to estimate the binding energy of the ejected envelope one needs to calculate the difference between the final and initial total binding energies of the donor star \citep[see Eq. 62 of][]{2010ApJ...717..724G}.
This is a reasonable approach and has been, e.g., used in the literature to study the progenitor channel of envelope-stripped Type Ib supernova iPTF13bvn \citep{2017MNRAS.466.3775H,2017MNRAS.469L..94H}.
The binding energy we estimate in our study would not be directly comparable with this alternative definition of the binding energy.

\subsection{Implications for massive binary evolution}
\label{sub:binaries}
\subsubsection{Survival of the \ac{CE}}
Our estimates of $f_R$ suggest that (arguably) all of the systems we consider will experience Roche-lobe overflow immediately or shortly after the star has been stripped.
This naively implies that, for all our critical points and under our assumptions, none of the stars are able to completely eject the envelope.
\cite{FragosMesa} used one-dimensional stellar evolution to emulate the \ac{CE} phase of a massive star with a \ac{NS} companion and concluded that super-efficient ($\alpha>1$) energy sources are needed to eject the envelope.
This is in broad agreement with our results.

\cite{Klencki2021A&A} argued that there is usually not enough energy to expel the envelope unless it is deeply convective. In the case of systems with \ac{NS} accretors, they found successful \ac{CE} ejection only when the donor initiated the \ac{CE} phase already after central helium exhaustion \citep[see Fig. B.2 from][]{Klencki2021A&A}, aka ``case C" mass transfer. This is consistent with our findings of $f_R<1$ in all the considered cases, irrespective of the choice of bifurcation point.
Notably, the case C donors in \cite{Klencki2021A&A} had envelopes that were more deeply (nearly fully) convective, making their binding energies lower compared to the convective-envelope Hertzsprung gap donors considered in this study.

Finally, we highlight and reinforce that stripping and post-stripping evolution can result in significant differences in the fate of post-\ac{CE} binaries. 
Tables \ref{tab:lambdaSolar}, \ref{tab:lambdaMW}, and \ref{tab:lambdaLMC} show that a monotonically increasing value for the remaining envelope mass does not always results in a monotonically increasing value of $f_R$.

\subsubsection{Hydrogen abundance of stripped stars}
Throughout this paper, we have presented and discussed the uncertainties in the remaining envelope mass in stripped stars.
However, there is an additional observable property that is correlated with the remaining envelope mass: the surface hydrogen mass fraction (see Fig. \ref{fig:StarProfile12Zsun}).

\cite{2018A&A...611A..75S} numerically explored the hydrogen-shell abundance gradient between the core and the envelope of stripped stars, in the context of apparently-single and binary systems, and compared them with the Wolf-Rayet population in the Small Magellanic Cloud.
Most of their models assume that  the slope of the hydrogen gradient is steep, and that the surface hydrogen mass fraction is $\lesssim 0.3-0.4$ \citep[see Fig. A.1 and A.2 from][]{2018A&A...611A..75S}.

Our 12 $\rm{M_{\odot}}$ model at $Z=Z_{\odot}$ (Fig. \ref{fig:StarProfile12Zsun}) shows the hydrogen mass fraction of that particular stellar model, which is shallower than most synthetic gradients from \cite{2018A&A...611A..75S}.
Stripping that star until the upper bifurcation point results in $M_{\rm{env,rem}}=0.95\ \rm{M_{\odot}}$ above the 2.71 $\rm{M_{\odot}}$ core, leading to a hydrogen mass fraction $<0.2$. 
For the other (deeper) critical points, the hydrogen mass fraction would be even lower.
While all of our models are at higher metallicity than that of the Small Magellanic Cloud, and therefore cannot be directly compared with the sample and models from \cite{2018A&A...611A..75S}, we believe their method can be useful to further constrain the progenitor properties of stripped stars.

\citet{2020MNRAS.495.4659F} systematically studied the connections between internal and surface properties of massive stars at $Z=0.02$.
They compute and provide surface properties for models with different composition, core mass, and remaining envelope mass.
They consider less massive stars than \citet{2018A&A...611A..75S}, which are more comparable to our models.
In particular, they present the luminosity and effective temperature of stripped stars.
For models with core mass $<8.4\ \rm{M_{\odot}}$ and $M_{\rm{env,rem}}=1.2\ \rm{M_{\odot}}$, such as our $Z=Z_{\odot}$ models (Table \ref{tab:lambdaSolar}), they predict that the uncertainty in luminosity is $\lesssim 0.2$ dex, and the uncertainty in the effective temperature is $\lesssim 0.1$ dex \citep[see Fig. 13 of ][]{2020MNRAS.495.4659F}.
For their stripped-star case study HD 45166 \citep{2005A&A...444..895S} they estimated a core mass of $2.30^{+0.35}_{-0.23}\ \rm{M_{\odot}}$, where they define the core as $X_H=10^{-4}$, and an envelope mass of $0.15^{+0.11}_{-0.08}\ \rm{M_{\odot}}$.
Directly comparing these results to our models, it seems like this particular binary would have experienced stripping deep into its envelope, likely close to the maximum compression point.
However, while HD 45166 is a system of interest, there are many relevant observational  \citep{2019ASPC..519..197D} and modelling \citep{2017A&A...608A..11G,GotbergSES,2020MNRAS.495.4659F} uncertainties.

\cite{2021arXiv211110271K} recently proposed stripped stars with larger remaining envelope masses as products of mass transfer evolution in low-metallicity massive binaries. They showed that donors for which a sufficiently large fraction of the envelope is left unstripped will have much lower effective temperatures compared to classical stripped stars. Such partially-stripped stars could at a first glance appear as normal B-type stars, as was the case with LB-1 and HR 6819 systems \citep{2019Natur.575..618L,2020A&A...639L...6S,2020A&A...641A..43B,2020MNRAS.495.2786E}.

\subsection{Three-dimensional hydrodynamic simulations}
\label{3dsimulations}
Recently, pioneering three-dimensional hydrodynamic simulations of the common-envelope phase of solar-metallicity massive stars with \ac{NS} companions have appeared in the literature. 

\cite{2020arXiv201106630L} and \cite{2021arXiv211112112M} performed mesh simulations of the \ac{CE} phase with a massive donor and a 1.4 $\rm{M_{\odot}}$ \ac{NS} companion. 
\cite{2020arXiv201106630L} performed adaptive-mesh refinement simulations of an initially 12 $\rm{M_{\odot}}$ stellar model which becomes very extended, reaching radii larger than 1000 $\rm{R_{\odot}}$.  However, their hydrodynamic modeling did not follow the early evolution of the \ac{CE} phase.
They estimate values of $\alpha \approx 0.1-0.4$, but this estimate does not include internal energy in the calculation of the binding energy.
\cite{2021arXiv211112112M} performed a magnetohydrodynamical moving-mesh simulation of an initially 10 $\rm{M_{\odot}}$ stellar model that has a radius of 438 $\rm{R_{\odot}}$ at the beginning of the simulation.
They follow the evolution until the orbital separation stalls at $\approx 15\ \rm{R_{\odot}}$.
They estimate values of $\alpha \approx 0.61-2.29$, depending on. whether or not they include internal energy and on the location of the core/envelope boundary.

\cite{2022MNRAS.tmp...68L} performed smoothed-particle hydrodynamic simulations of the \ac{CE} phase with a 1.26 $\rm{M_{\odot}}$ \ac{NS} companion.
They simulated an initially 12 $\rm{M_{\odot}}$ donor that has a radius of 619 $\rm{R_{\odot}}$ at the beginning of the simulation.
In this simulation, the \ac{NS} companion has not stalled when it reaches the base of the convective envelope.
They do not simulate the inspiral into the hydrogen-shell region and don't quote values for $\alpha$ in the simulation with a \ac{NS} companion (but they do for the simulation with a 3 $\rm{M_{\odot}}$ black-hole companion, where the inspiral stalls).

While these results are not easy to directly compare with one another, these methods and papers are very promising steps in the progress toward a complete solution to the \ac{CE} phase in massive stars \citep[see also][]{2019IAUS..346..449R}.

\subsubsection{Additional hydrogen-rich mass transfer: case BA}
Our method of modelling \ac{CE} stripping indicates that will be an initial short ($t < 100$ yr) relaxation phase when all models expand. 
Longer-term ($t > 100$ yr) evolution and expansion depends on the amount of envelope mass retained by the stripped star.
Depending on the post-\ac{CE} orbital configuration, some expanding stars might promptly engage in a further mass transfer episode.
Moreover, some remnants with masses above the maximum compression point often have a second peak in their radial expansion when the remaining hydrogen in the envelope burns (Sec.~\ref{sub:radius}).
The overall trend is that, the more envelope mass is left, the greater will the overall expansion be.

\cite{Quast} considered stable mass transfer on a nuclear timescale in high-mass X-ray binaries. 
In that study, which considers more massive helium cores than the ones presented here, the timescale is related to the time span of core helium burning.   This would last for $\lesssim 10^6$ years for the donor stars considered here. 
If the expansion of the remaining envelope leads to mass transfer, and if that mass transfer is stable and occurs on a nuclear timescale, then the \ac{NS} companion could in principle accrete some of this transferred mass.
If we assume the Eddington limit as the mass accretion rate, the \ac{NS} could accrete up to $\lesssim 10^{-2}$ $\rm M_{\odot}$.
This amount of mass is non-neglibile for a \ac{NS}, as it could lead to spin-up and (mild) pulsar recycling \citep[see, e.g.,][and references therein]{TaurisPulsar,TaurisFormation}.
The early post-\ac{CE} mass transfer phase can also change the orbital separation of the binary.  
Since the \ac{NS} companion is less massive than the stripped remnant, it is likely the orbital separation will decrease \citep[e.g.,][]{FragosMesa}. 
The orbital evolution of this case BA mass transfer phase after a \ac{CE} can result in an prolonged mass transfer episode.

\subsection{Conclusions}
\label{sub:conclusions}
We used one-dimensional single stellar evolution methods to explore the \ac{CE} phase of massive binaries that may become \acp{BNS}.
We did this for a range of donor masses and metallicities.
We focused on stellar evolution after stripping during a \ac{CE} episode, particularly on the radial evolution of stripped stars as a function of the remaining envelope mass.
We explored how deeply a star can be stripped without experiencing Roche lobe overflow immediately after the \ac{CE} phase. We considered the bifurcation points between  envelope contraction and prompt re-expansion as boundaries between the retained core and ejected envelope. We found that these bifurcation points lie above the maximum compression points, which are commonly used as the location of the core/envelope boundary. 
This implies that the \ac{CE} phase could stall at larger radii than generally thought.
Consequently, post-\ac{CE} donors may still retain 1 to 2 M$_\odot$ of a hydrogen-rich envelope in our models.
Finally, if we consider orbital energy to be 100\% efficient in unbinding the envelope whose binding energy includes gravitational, thermal, radiation and recombination energies, then all of our models would overfill their Roche lobe shortly after ejecting the envelope.

\section*{Acknowledgments}
We thank Ryosuke Hirai, Matthias Kruckow, Abel Schootemeijer, and the anonymous referee for useful discussions and suggestions.
AVG acknowledges support by the Danish National Research Foundation (DNRF132).
JK, AI and GN acknowledge support from the Dutch Science Foundation NWO. IM is a recipient of the Australian Research Council Future Fellowship FT190100574 and acknowledges support from the Australian Research Council Centre of Excellence for Gravitational Wave Discovery (OzGrav), through project number CE17010000. 

\section*{Data Availability}
Main data are incorporated into the article.
MESA inlists are available in a repository and can be accessed via \doi{10.5281/zenodo.5155790} \citep{vigna_gomez_alejandro_2021_5155790}.

\bibliographystyle{mnras}
\bibliography{References}

\label{lastpage}
\end{document}